\documentclass[twocolumn,reprint,nofootinbib,longbibliography,prd]{revtex4-1}
\usepackage{float}
\usepackage{tikz,pgfplots}
\usepackage{subfigure}
\usepackage{amsfonts, amsmath, amssymb, bm, enumerate, graphicx, graphics, color,mathrsfs,hyperref,nicefrac}
\usepackage[utf8]{inputenc}

\newcommand{\Lie}[0]{{\cal L}\, }

\newcommand{\td}{\text{d}}
\newcommand{\tl}{\theta_{(\ell)}}

\newcommand{\be}{\begin{equation}}
\newcommand{\ee}{\end{equation}}
\newcommand{\bea}{\begin{eqnarray}}
\newcommand{\eea}{\end{eqnarray}}

\newcommand{\Heq}{\overset{\scriptscriptstyle{H}}{=}}

\newcommand{\eqq}[1]{\overset{\scriptscriptstyle{#1}}{=}}

\newcommand{\cV}{\mathcal{V}}

\begin{document}

\title{	
Horizons as boundary conditions in spherical symmetry}

\author{Sharmila Gunasekaran}
\email{sdgg82@mun.ca}
\author {Ivan Booth}
\email{ibooth@mun.ca} 
%\affiliation{
%Department of Mathematics and Statistics\\ Memorial University of Newfoundland \\  
%St.~\!\!John's, Newfoundland and Labrador, A1C 5S7, Canada
%}
\affiliation{
Department of Mathematics and Statistics\\ Memorial University of Newfoundland \\  
St.~\!\!John's, Newfoundland and Labrador, A1C 5S7, Canada \\
}
\date{\today}

\begin{abstract}
We initiate the development of a horizon-based initial (or rather final) value formalism to describe the  geometry and physics of the near-horizon spacetime: 
data specified on the horizon and a future ingoing null boundary determine the near-horizon geometry. In this initial paper we restrict our attention to spherically symmetric spacetimes made dynamic
by matter fields. We illustrate the formalism by considering a  black hole interacting with a) inward-falling, null matter (with no outward flux) and b) a massless scalar field. The inward-falling
case can be exactly solved from horizon data.  For the more involved case of the scalar field 
we analytically investigate the near slowly evolving  
horizon regime and  propose a numerical integration for the general case. 
%We also discuss the case of null dust and timelike dust in the appendix to illustrate the horizon-based approach used here.   
\end{abstract}

\maketitle

\section{Introduction}

%In this paper 
This paper begins an investigation into what  horizon dynamics can tell us about external black hole physics. At first thought this might seem obvious: if one watches a numerical
simulation of a black hole merger and sees a post-merger horizon ringdown (see for example \cite{sxs}) then it is natural to think of that oscillation as a source of emitted gravitational waves.
However % on second thought 
 this  cannot be the case. Neither event nor apparent horizons can actually send signals to infinity: apparent horizons lie inside event horizons which
in turn are the boundary for signals that can reach infinity\cite{Hawking:1973uf}. It is not horizons themselves that interact but rather the ``near horizon'' fields. This idea was (partially) formalized
 as  a ``stretched horizon''  in the membrane paradigm\cite{Thorne:1986iy}. 

Then the best  that we can hope for from horizons is that they act as a proxy for the near horizon fields with horizon evolution reflecting some aspects of their dynamics. 
As explored in \cite{Jaramillo:2011re,Jaramillo:2011rf,Jaramillo:2012rr,Rezzolla:2010df,Gupta:2018znn} there should then be a correlation between horizon evolution and 
external, observable, black hole physics. 

Robinson-Trautman spacetimes (see for example \cite{Griffiths:2009dfa}) demonstrate that this correlation cannot  be perfect. In those spacetimes there can be 
outgoing gravitational (or other) radiation arbitrarily close to an isolated (equilibrium) horizon\cite{Ashtekar:2000sz}. 
Hence our goal is  two-fold: both to understand the conditions under which a correlation will exist and to learn precisely what information it contains.

The idea that horizons should encode physical information about black hole physics is not new. The classical definition of a black hole as the complement of the causal past of future null infinity \cite{Hawking:1973uf} is essentially global and so defines a black hole spacetime 
 rather than a black hole \emph{in} some spacetime. However there
are also a range of geometrically defined black hole boundaries based on outer and/or marginally trapped surfaces that seek to localize black holes. These include 
apparent\cite{Hawking:1973uf}, trapping\cite{Hayward:1993wb}, isolated \cite{Ashtekar:1998sp,Ashtekar:1999yj,Ashtekar:2000sz,PhysRevD.49.6467} and dynamical 
\cite{Ashtekar:2003hk} horizons as well as
future holographic screens \cite{Bousso:2015qqa}. These quasilocal definitions of black holes have successfully localized black hole mechanics to the 
horizon\cite{Ashtekar:1998sp,Ashtekar:1999yj,Ashtekar:2003hk,Booth:2003ji,Bousso:2015qqa,Hayward:1993wb} and been particularly 
useful in formalizing what it means for a (localized) black hole to evolve or be in equilibrium. They are used in numerical relativity  not only as excision surfaces (see, for example 
the discussions in \cite{Baumgarte:2010ndz,Thornburg:2006zb}) but also in
interpreting physics (for example \cite{Dreyer:2002mx,Cook:2007wr,Chu:2010yu,Jaramillo:2011re,Jaramillo:2011rf,Jaramillo:2012rr,Rezzolla:2010df,Lovelace:2014twa,Gupta:2018znn,Owen:2017yaj}).
%They are also much studied in mathematical relativity where, for example, they have been used in the 
%classification of possible black hole solutions \cite{Kunduri:2013ana} and establishing inequalities that restrict the possible physical properties of solutions 
%\cite{Dain:2006wb,Jaramillo:2011pg}. 

%Studying the neighborhood of the horizon is also relevant to classifying near horizon geometries and addressing the uniqueness problems for black holes \cite{Li:2015wsa}. The approach of using horizons to understand it's neighborhood is particularly relevant in the study of isolated horizons. Since the horizon is characterisitic in this case, the approrpiate way to study the system is as a characteristic initial value problem \cite{MR1032984,doi:10.1063/1.1724305,Krishnan:2012bt}. See for instance\cite{Krishnan:2012bt}, where data is specified on the horizon and an outgoing past null cone. 

%Studying the neighborhood of the horizon is also relevant to classifying near horizon geometries and addressing the uniqueness problems for black holes \cite{Li:2015wsa}.

\begin{figure}
\centering     %%% not \center
\subfigure[ Future and past domains of dependence for ${\cal H}_{\text{dynamic}}$: standard (3+1) IVP ]{\label{hd1}\includegraphics[width=65mm]{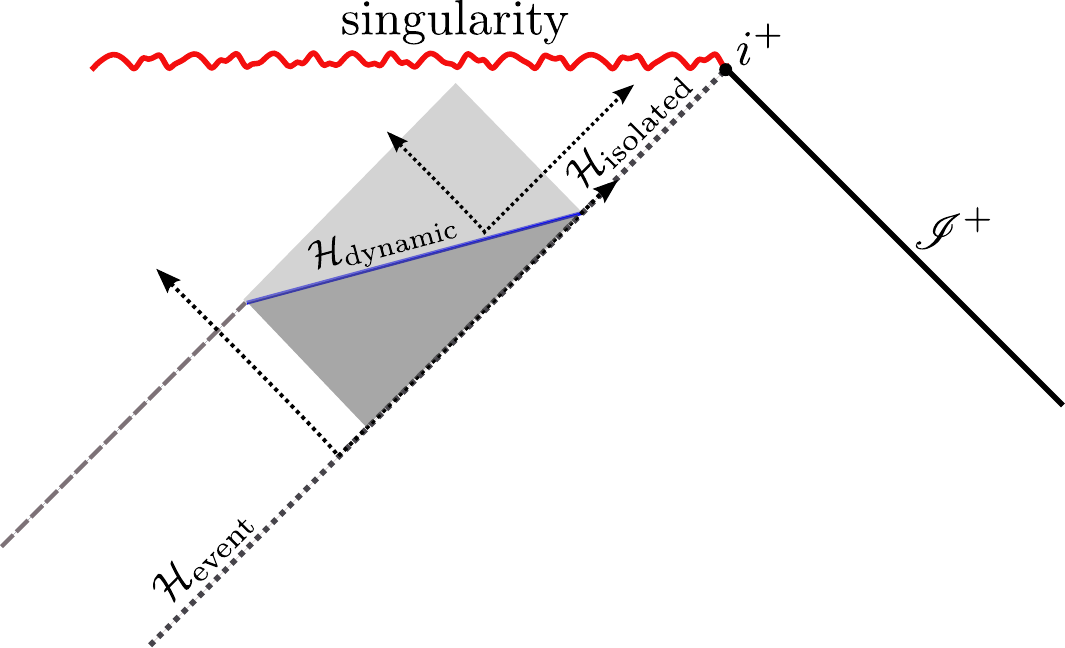}}
\subfigure[ Future and past domains of dependence for ${\cal H}_{\text{dynamic}} \cup {\cal N} $: characteristic IVP ]{\label{hd2}\includegraphics[width=65mm]{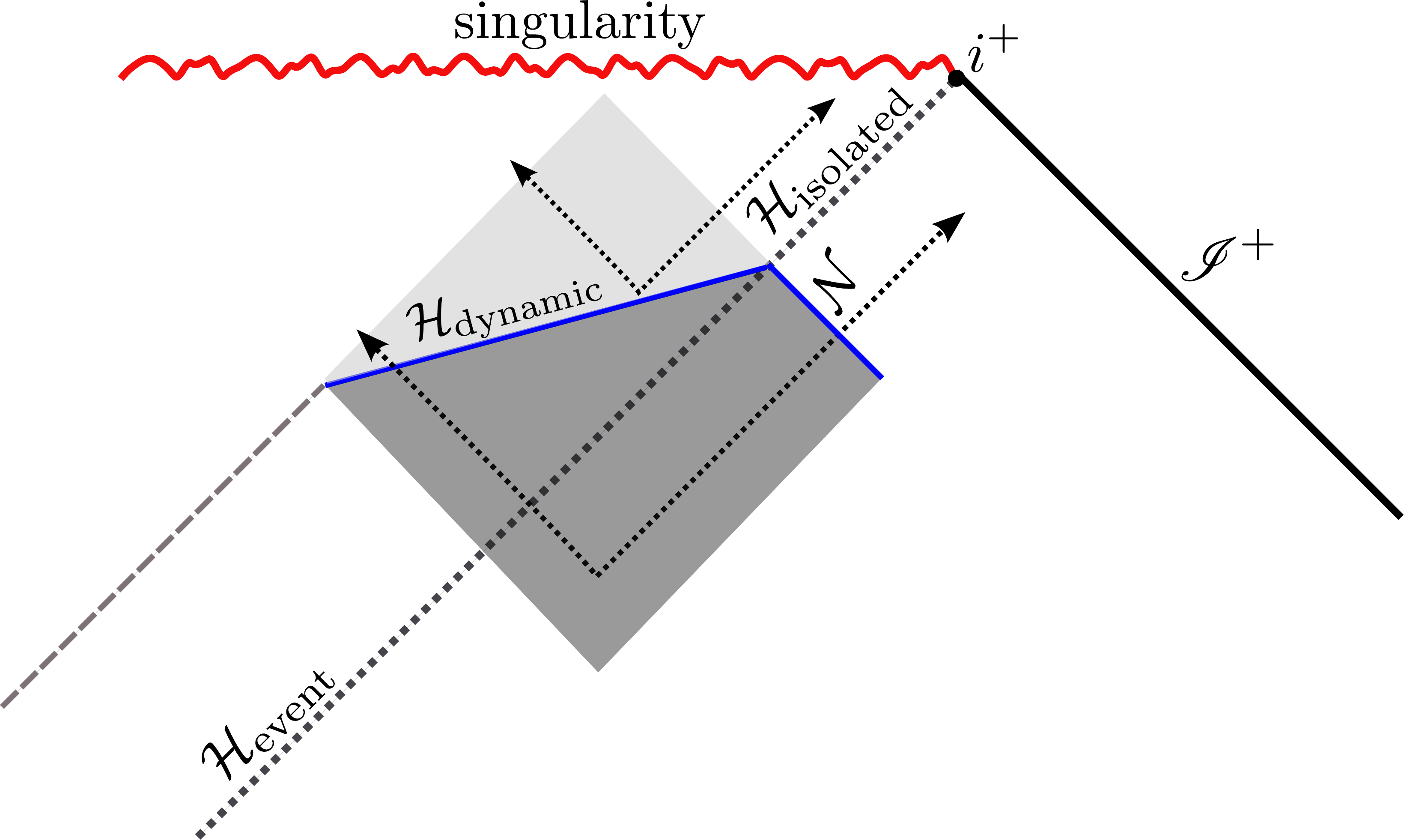}}
\caption{Domains of dependence of initial data}
\label{dod}
\end{figure}
In this paper we work to quantitatively link horizon dynamics to observable black hole physics. To establish an initial  framework and build intuition we for now restrict our
 attention to spherically symmetric marginally outer trapped tubes (MOTTs) in similarly symmetric spacetimes. Matter fields are included to drive the dynamics. Our primary  approach is to take 
horizon data as a (partial) final boundary condition that is used to determine the fields in a region of spacetime in its causal past. In particular these boundary conditions constrain
the geometry and physics of the associated ``near horizon'' spacetime. The main application that we have in mind is interpreting the physics of evolving horizons  that have been generated by 
either numerical simulations or theoretical considerations. 

Normally, data on a MOTT by itself is not sufficient to specify any region of the external spacetime. As shown in FIG.~\ref{hd1} even for a spacelike MOTT (a dynamical horizon) 
the region determined by a standard (3+1) initial value formulation would lie entirely within the event horizon. More information is needed to determine the near-horizon spacetime
and hence in this paper we work with a characteristic initial value formulation  \cite{MR1032984,doi:10.1063/1.1724305,Winicour:2012znc,Winicour:2013gha,Madler:2016xju} where extra data is specified on a null surface 
$\mathcal{N}$ that is transverse to the horizon (FIG.~\ref{hd2}). Intuitively the horizon records inward-moving information while $\mathcal{N}$ records the outward-moving information. Together they are sufficient to reconstruct the spacetime.

There is an existing literature that studies spacetime near horizons, however it does not exactly address this problem. Most works focus on isolated horizons.  
\cite{Li:2015wsa} and \cite{Li:2018knr} examine spacetime near an isolated extremal horizon as a Taylor series expansion of the horizon 
 while \cite{Krishnan:2012bt} and  \cite{Lewandowski:2018khe} study spacetime near more
general isolated horizons but in a characteristic initial value formulation with the extra information specified on a transverse null surface. \cite{Booth:2012xm} studied both the isolated
and dynamical case though again as a Taylor series expansion off the horizon. In the case of the Taylor expansions, as one goes to higher and higher orders one needs to know higher and higher order derivatives of metric quantities at the horizon to continue the expansion. While
the current paper instead investigates the problem as a final value problem,  it otherwise closely follows the notation of and uses many results from \cite{Booth:2012xm}.

It  is organized as follows. We introduce the final value formulation of spherically symmetric general relativity in Sec.\ref{sec:formulation}. We illustrate this for infalling null matter in \ref{sec:matter modelsI} and then the much more interesting massless scalar field 
in Sec.\ref{sec:matter modelsII}. We conclude with a discussion of results in Sec.\ref{sec:discussion}.

\section{Formulation}
\label{sec:formulation}

\subsection{Coordinates and metric}

We work in  a  spherically symmetric spacetime (${\cal M}, g$) and a coordinate system whose non-angular coordinates
are $\rho$ (an ingoing affine parameter) and $v$ (which labels the ingoing null hypersurfaces and increases into the future).
%, $S_v$. 
Hence, $g_{\rho \rho} =0$ and the curves tangent to the future-oriented inward-pointing
\begin{equation}
N =  \frac{\partial}{\partial \rho}  \label{E1}
\end{equation}
 are null. We then scale $v$ so that $\mathcal{V}= \frac{\partial}{\partial v}$ satisfies
 \begin{equation}
 \mathcal{V} \cdot N = -1. \label{E2}
 \end{equation}
One coordinate freedom still remains: the scaling of the affine parameter on the individual null geodesics 
\begin{equation}
\tilde{\rho} = f(v) \rho \, . \label{gaugefree}
\end{equation}
 In subsection \ref{physconf}  
we will fix this freedom by specifying how $N$ is to be scaled along the $\rho = 0$ surface $\Sigma$ (which we
 take to be a black hole horizon).
%
%black hole horizon (which will be taken to be the $\rho = 0$ surface). 
\begin{figure}[h]
\begin{center}
\includegraphics[scale=0.25]{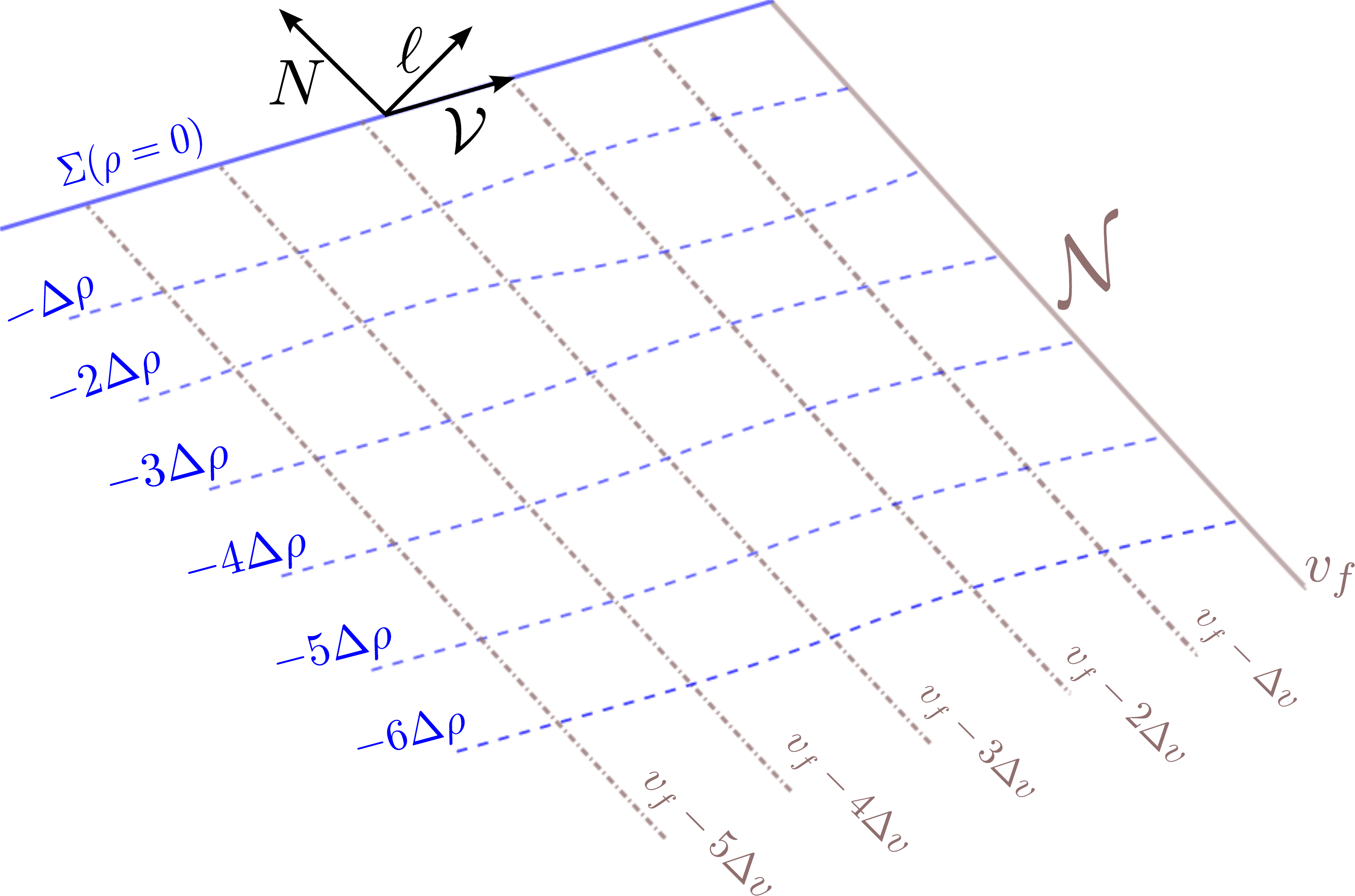}
\caption{Coordinate system for characteristic evolution. We work with final boundary conditions so that in the region of interest $\rho < 0$.  }
\label{fig:3p1}
\end{center}
\end{figure}

Next we define the future-oriented outward-pointing null normal to the spherical surfaces $S_{(v,\rho)}$ as $\ell^a$ and scale  so that 
\begin{align}
\ell \cdot  N = -1 \label{affine2} \, . 
\end{align}
With this choice  the four-metric $g_{ab}$ and induced two-metric $\tilde{q}_{ab}$ on the $S_{(v,\rho)}$ are related by
\begin{equation}
g^{ab} = \tilde{q}^{ab} - \ell^a N^b - N^a \ell^b \, .
\label{decomp}  
\end{equation}
Further for some function $C$ we can write 
\begin{equation}
{\cal V} = \ell - C N \, . \label{crel1}
\end{equation}
The coordinates and normal vectors are depicted in FIG.\ref{fig:3p1}
%Future pointing null normals  span the normal space to $S_v$: we label the outward and inward pointing vectors as $\ell^a (ou \text{ and } N^a$) respectively. These are defined up to a choice of reparametrization. We choose a scaling so that,
%\begin{align}
%\ell \cdot  N = -1 \,, \quad \mathcal{V} \cdot N = -1 \label{affine2}
%\end{align}
%where, $ {\cal V} =  \frac{\partial}{\partial v}$ . The coordinate set up is best described through Fig.\ref{fig:3p1}.
%The affine freedom in $\rho$ is fixed by making  $\rho$ an off-horizon coordinate and \eqref{affine2}. The foliation is depicted below in Fig.\ref{fig:3p1}.
and give the following form of the metric: 
\begin{equation}
ds^2 = 2 C(v,\rho) \td v^2 - 2 \td v \, \td \rho + R(v,\rho)^2 \td \Omega^2
\label{metant} 
\end{equation}
where $R(v,\rho)$ is the areal radius of the $S_{(v,\rho)}$ surfaces. Note the similarity to ingoing Eddington-Finkelstein 
coordinates for a Schwarzschild black hole. However $\nicefrac{\partial}{\partial \rho}$ points inwards as opposed to the 
outward oriented $\nicefrac{\partial}{\partial r}$ in those coordinates (hence the negative sign on the $\mbox{d} v \mbox{d} \rho$ cross-term). 
%The full metric and induced metric on the $S_{(v,\rho)}$ are related by the null normals:
%\begin{equation}
%g^{ab} = \tilde{q}^{ab} - \ell^a N^b - N^a \ell^b.
%\label{decomp}  
%\end{equation}
%Now, ${\cal V}, \ell $ and $N$ can be related by
%\begin{equation}
%{\cal V}^a = \ell^a - C N^a
%\label{crel1}
%\end{equation}

Typically, as shown in FIG.\ref{fig:3p1} we will be interested in regions of spacetime that are bordered in the future by a surface $\Sigma$ of 
indeterminate sign on which $\rho=0$ and a null  $\mathcal{N}$ which is one of the $v$=constant surfaces (and so 
$\rho < 0$ in the region of interest). We will explore how data on 
those surfaces determines the region of spacetime in their causal past. 
 
\subsection{Equations of motion}
 In this section we break up the Einstein equations relative to these coordinates, beginning by defining some geometric quantities that
appear in the equations. 
 
First  the null expansions for the $\ell^a$ and $N^a$ congruences are
 \begin{align}
 \theta_{(\ell)} &= \tilde{q}^{ab} \nabla_a \ell_b = \frac{2}{R} \Lie_\ell R \quad \mbox{and}  \label{nullexp1}  \\
 \theta_{(N)} &= \tilde{q}^{ab} \nabla_a N_b  = \frac{2}{R} \Lie_N R = \frac{2}{R} R_{,\rho}. 
 \label{nullexp2} 
 \end{align}
 while the inaffinities of the null vector fields are 
 \begin{align}
 \kappa_N &= - N^a N_b \nabla_a \ell^b = 0 \quad \mbox{and} \\
 \kappa_\mathcal{V} &= \kappa_\ell - C \kappa_N  =  - \ell^a N_b \nabla_a \ell^b \,. 
 \end{align} 
By construction $\kappa_N = 0$ and so we can drop it from our equations and henceforth write 
 \begin{equation}
 \kappa \equiv \kappa_{\mathcal{V}} = \kappa_\ell \, . 
 \end{equation}
%Geometrically $\kappa$  may also be understood as the connection in the $(\ell,N)$-plane along curves of constant $\rho$ that 
%rescales a derivatives in the $\ell$  into its affinely parameterized form:
%\begin{equation}
%\left( \ell^a \nabla_a + \kappa \right) \ell^b = 0 \, . 
%\end{equation}
 Finally the Gaussian curvature of $S_{(v,\rho)}$ is:
\begin{equation}
\tilde{K} = \frac{1}{R^2} \, . 
\end{equation}

Then these curvature quantities are related by constraint equations along the surfaces of constant $\rho$ 
\begin{align}
\mathcal{L}_\cV  R = & \Lie_\ell R - C \Lie_N R  \quad \text{(by definition)}\, , 
\label{const1} \\
\mathcal{L}_\mathcal{V} \theta_{(\ell)} = &  \kappa  \, \theta_{(\ell)} + C \left(  \frac{1}{R^2 } + \theta_{(N)} \theta_{(\ell)} - G_{ \ell N}  \right) \nonumber \\ & - \left(  G_{\ell \ell}  + \frac{1}{2} \theta_{(\ell)}^2 \right) \,, 
\label{const2}\\
\mathcal{L}_\mathcal{V} \theta_{(N)} = &  -\kappa\,  \theta_{(N)} -  \left( \frac{1}{R^2 } + \theta_{(N)} \theta_{(\ell)} - G_{\ell N}  \right) \nonumber \\  & + C \left(  G_{\! N \! N}  + \frac{1}{2} \theta_{(N)}^2 \right), 
\label{const3}
\end{align}
and ``time'' derivatives in the $\rho$ direction 
\begin{align}
{\cal{L}}_{N} \theta_{(N)} = & -\frac{\theta_{(N)}^2}{2} - G_{\!N \!N},  \label{cevol1} \\ 
{\cal{L}}_N \theta_{(\ell)} =& -\frac{1}{R^2} - \theta_{(N)} \theta_{(\ell)} + G_{\ell N},  \label{cevol2} \\
{\cal{L}}_N \kappa = & \frac{1}{R^2 } + \frac{1}{2} \theta_{(N)} \theta_{(\ell)} - \frac{1}{2}  G_{\tilde{q}}  -   G_{\ell N},  \label{cevol3}
\end{align}
where by the choice of the coordinates
\begin{equation}
\kappa =  {\cal{L}}_N C  \, .  \\
\end{equation}
These equations can be derived from the variations for the corresponding geometric quantities (see, for example, \cite{Booth:2006bn} and \cite{Booth:2012xm}) and of course are coupled to the matter content of the system through
the Einstein equations 
\begin{equation}
G_{ab } = 8 \pi T_{ab} \; . 
\end{equation}

Using (\ref{nullexp1}) and (\ref{nullexp2})  we can rewrite the constraint and evolution equations in terms of the metric coefficients and coordinates as:
%{\bf plus $\ell N$ terms}
\begin{align}
 R_{,v} = & \, R_\ell - C R_N \,, \quad 
\label{const1a} \\
 R_{\ell,v} = & \,  \kappa R_{\ell} +   \frac{C \left( 1+ 4 R_\ell R_N   \right)}{2 R}  - \frac{R}{2} \,( G_{\ell \ell} + C G_{\ell N}) \,, 
\label{const2a}\\
 R_{N,v} = & -\kappa R_{N} -    \frac{ \left( 1+ 4 R_\ell R_N   \right)}{2 R}  + \frac{R}{2} \, (  G_{\ell N}  + CG_{N\! N}).
\label{const3a}
\end{align}
and
\begin{align}
R_{,\rho \rho}  &= - \frac{R}{2} \, G_{\! N \! N} \label{fcevol1}, \\ 
({R R_{\ell}})_{,\rho} &= -\frac{1}{2} + \frac{R^2}{2} G_{\ell N}, \label{fcevol2} \\
 C_{,\rho \rho} &= \frac{1}{R^2} + \frac{2R_{\ell} R_{N }}{R^2} -  \frac{1}{2}  G_{\tilde{q}} - G_{\ell N}, \label{fcevol3}
\end{align}
where 
\begin{equation}
\kappa  = C_{,\rho}  \label{fcevol0}. \\
\end{equation}
%\paragraph*{Constraint equations}
%\begin{align}
%\mathcal{L}_\mathcal{V} R = & R_\ell - C R_N \,, \quad 
%\label{const1a} \\
%\mathcal{L}_\mathcal{V} R_{\ell} = &  \kappa_\mathcal{V} R_{\ell} +   \frac{C \left( 1+ 2 R_\ell R_N   \right)}{2 R}  - \frac{R}{2}G_{\ell \ell}  \,, \text{ and }
%\label{const2a}\\
%\mathcal{L}_\mathcal{V} R_{N} = & - \kappa_\mathcal{V} R_{N} -    \frac{ \left( 1+ 2 R_\ell R_N   \right)}{2 R}  + \frac{C R}{2}G_{\! N \! N} \, .
%\label{const3a}
%\end{align}
For those who don't want to work through the derivations of  \cite{Booth:2006bn} and \cite{Booth:2012xm},  these can also be derived fairly easily (thanks to the spherical symmetry)
from an explicit calculation of the Einstein tensor for (\ref{metant}).

\subsection{Final Data}
\label{physconf}

%We now consider how these equations can be used to solve for regions of spacetime with initial/final data specified on a $\rho = 0$ surface $\Sigma$ along with 
%a $v=\mbox{constant}$ surface $\mathcal{N}$. 
%

We will focus on the case where $\rho =0$  is an isolated or dynamical horizon $H$. Thus 
 \begin{equation}
 \tl \Heq 0 \quad \Longleftrightarrow \quad  R_\ell \Heq  0 \; .   
 \end{equation}
The notation $\Heq$ indicates that the equality holds on $H$ (but not necessarily anywhere else). Further we can  use the coordinate freedom  (\ref{gaugefree}) to set
\begin{equation}
R_N \Heq  R_{,\rho}| \Heq -1 \, . \label{RIC}
\end{equation}

On $H$,  the constraints (\ref{const1a})-(\ref{const3a}) fix three of 
\begin{align}
\{C, \kappa, R, R_\ell, R_N, G_{\ell \ell}, G_{\ell N}, G_{NN}  \}
\end{align}
given the other five quantities. For example if $R_\ell \Heq 0$ and $R_N \Heq -1$ then 
 (\ref{const1a})  and (\ref{const2a}) give
\begin{equation}
R_{,v} \Heq C \Heq \frac{R^2 G_{\ell \ell}}{1 -  R^2 G_{\ell N}}
  \label{Rp}
\end{equation}
%or 
%\begin{equation}
%C_H = \frac{R_H^2 G_{\ell \ell}|_H}{1 - C_H R_H^2 G_{\ell N}|_H}
%\end{equation}
and (\ref{const3a}) gives
\begin{equation}
\kappa = C_{\rho} \Heq   \frac{1}{2R} -  \frac{R}{2} \left(G_{\ell N} +  C  G_{\! N \! N}  \right) \label{kappaH} \, . 
\end{equation}
Thus if $G_{\ell \ell}$ and $G_{\ell N}$ are specified for $v_i \leq v \leq v_f$ on $H$  and $R(v_f) \Heq R_f$ then  one can solve 
(\ref{Rp}) to find $R$ over the entire range.  
%\begin{equation}
%R_H= \frac{R_f}{1+R_f \int_v^{v_f} \! \! G_{\ell \ell}|_H dv} \, . 
%\end{equation}
Equivalently one could take $R$  and one of $G_{\ell \ell}$ or $G_{\ell N}$ as primary and then solve for the other component of the stress-energy. 

Of course, in general the matter terms will also be constrained by their own equations; these will be treated in later sections. Further data on $\rho = 0$ will generally not be 
sufficient to fully determine the regions of interest and data will also be needed on an $\mathcal{N}$. Again this will depend on the specific matter model. 

Nevertheless if there is a MOTT at $\rho = 0$ then the   constraints provide significant information about the horizon. 
If $G_{\ell \ell} = 0$ (no flux of matter through the horizon)  then we have an isolated horizon with $C=0$, a constant $R$ and a null $H$. This is independent of other 
components of the stress-energy. 

\begin{figure}
\centering     %%% not \center
\subfigure[$ $ Isolated horizon : $d\rho$ is timelike for all values of $\rho$ ]{\label{fol2}\includegraphics[width=70mm]{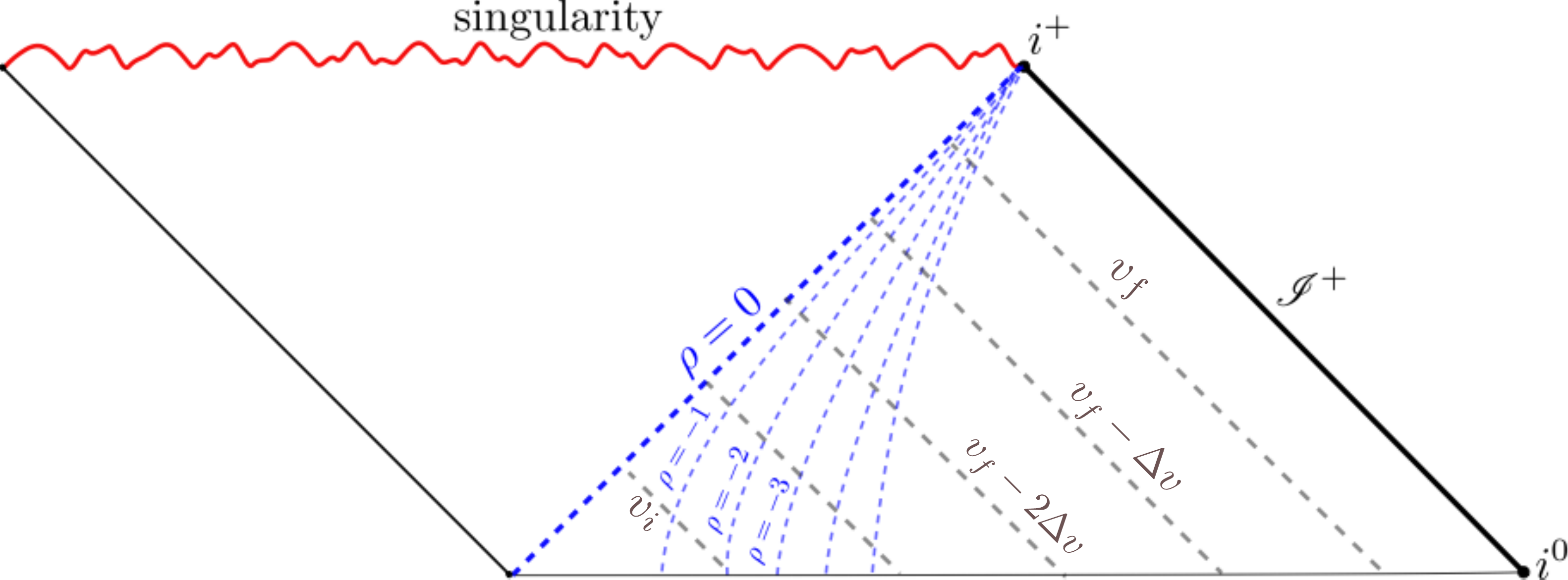}}
\subfigure[$ $  Dynamical horizon :  $d\rho$ is spacelike for small values of $\rho$ and eventually becomes timelike for large values of $\rho$]{\label{fol1}\includegraphics[width=70mm]{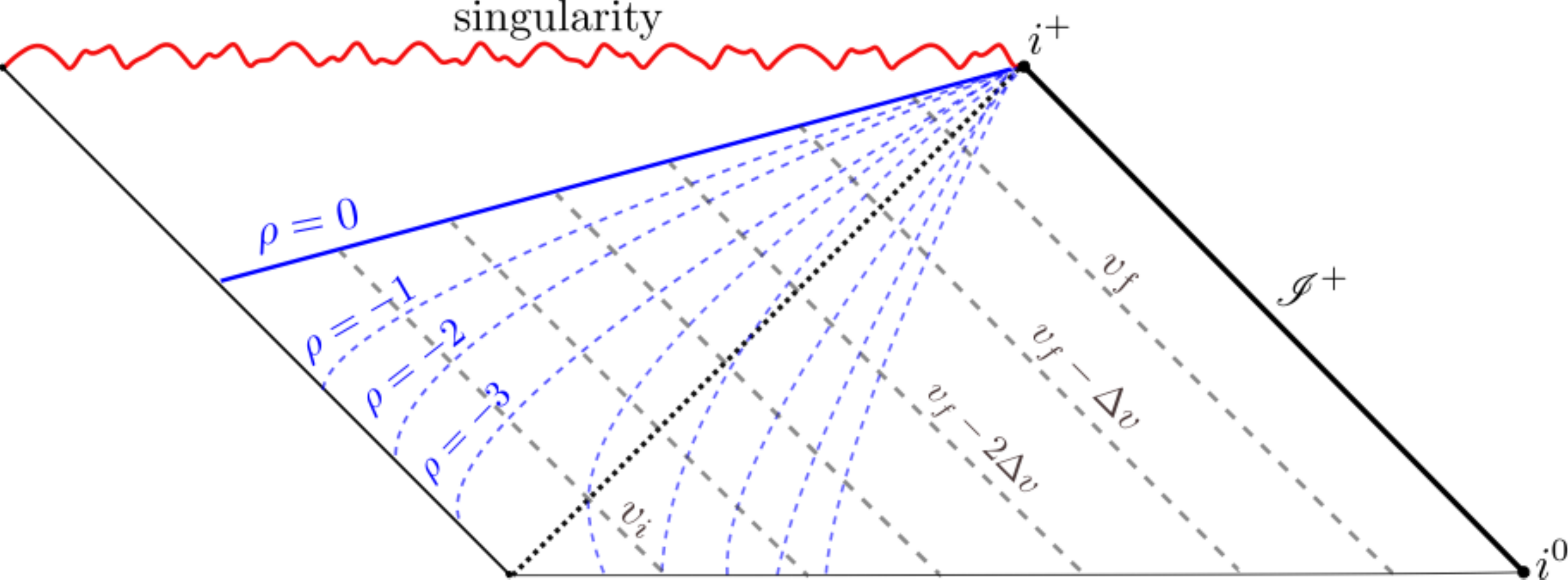}}
\caption{Spacetime foliation for isolated and dynamical horizons }
\label{dyniso}
\end{figure}

Alternatively if $G_{\ell \ell} > 0$ (the energy conditions forbid it to be negative) and $ G_{\ell N} < 1/R^2$ then we have a dynamical horizon with $C>0$, increasing $R$ 
and spacelike $H$\footnote{$G_{\ell N} > 1/R^2$ signals that another MOTS has formed outside the original one and so a numerical simulation would see  an apparent horizon ``jump'' \cite{Booth:2005ng,Bousso:2015qqa}. In the current paper all
matter satisfies $G_{\ell N} < 1/R^2$ and so  this situation does not arise. }. Note that this growth doesn't depend in any
way on $G_{\! N \! N}$: there is no sense in which the growing horizon ``catches'' outward moving matter and hence grows even faster. %
%
%then we have a dynamical horizon with $C>0$, $R$ growing and 
%$H$ spacelike. Note  that this expansion is still independent of the other components of the stress-energy: while the expanding horizon may ``grab'' more stress-energy
%this doesn't serve to further increase the expansion. However these other quantities do change the value of $\kappa_H$. 
%These  are special cases of the more general (non-spherical) cases consider in, for example, \cite{FOTH, MTT, IH, DH}. 
The behaviour of the coordinates relative to isolated and dynamical horizons along with $\mathscr{I}^+$ is illustrated in 
FIG.\ref{dyniso}. 

The evolution equations are more complicated and depend on the matter field equations. We examine two such cases in the following sections.

\section{Traceless inward flowing null matter}
\label{sec:matter modelsI}
As our first example consider matter that falls entirely in the inward  $N$-direction with no outward $\ell$-flux. Then data
on the horizon should be sufficient to entirely determine the region of spacetime traced by the horizon-crossing inward null geodesics: there are no dynamics that don't involve the horizon.

%
%
%
%We illustrate the formulation through two examples, a black hole interacting with a) traceless matter with no outward flux and b) massless scalar field.

Translating these words into equations, we assume that
\begin{equation}
T_{ab} N^a N^b = 0 \label{R1}
\end{equation}
(no matter flows in the outward $\ell$-direction).
Further, for simplicity we also assume that it is trace-free
\begin{equation}
g^{ab} T_{ab} = 0 \quad \Leftrightarrow \quad T_{\tilde{q}}  = 2 T_{\ell N} \; . \label{R2}
\end{equation}
Then we can solve for the metric using only the Bianchi identities
\begin{equation}
\nabla_a G^{ab} = 0, 
\end{equation}
without any reference to detailed equations of motion for the matter field. 
Keeping spherical symmetry but temporarily suspending the other  simplifying assumptions they may be written as:
\begin{align}
\mathcal{L}_\ell (R^2G_{\! N \! N}) + &  \mathcal{L}_N(R^2G_{\ell N})  + R^2( 2 \kappa_{\ell} G_{\! N \! N})  \nonumber \\ & + \frac{1}{2} R^2\theta_{(N)} G_{\tilde{q}} = 0,  \label{bianchi1} \\
\mathcal{L}_N(R^2G_{\ell \ell}) + & \mathcal{L}_\ell(R^2G_{\ell N})  + R^2(- 2 \kappa_{N} G_{\ell \ell}) \nonumber \\ & + \frac{1}{2} R^2 \theta_{(\ell)}  G_{\tilde{q}} = 0  \, .
\label{bianchi2}
\end{align}
In terms of metric coefficients with $\kappa_N = 0$ plus (\ref{R1}) and (\ref{R2}) these reduce to:
\begin{align}
(R^4 G_{\ell N})_{,\rho} = 0 \,\,\text{ and }  \label{constd1}\\ 
(R^2 G_{\ell \ell})_{,\rho} + \frac{1}{R^2} ( R^4 G_{\ell N} )_{,v} = 0.
\label{constd2}
\end{align}

As we shall see, this class of matter includes
interesting examples like Vaidya-Reissner-Nordstr\"om (charged null dust). 

We now demonstrate that given knowledge of $G_{\ell \ell}$ and $G_{\ell  N}$ over a region of horizon $\bar{H} = \{ H: v_i \leq v \leq v_f \}$ as well as $R(v_f) \Heq R_f$ we can 
determine the spacetime everywhere out along the horizon-crossing inward null geodesics.

\subsection{On the horizon}

First  consider the constraints on $\bar{H}$. In this case it is tidier to take $R$ and $G_{\ell N}$ as primary. Then we can specify
\begin{align}
R \Heq R_H (v) \quad \mbox{and} \quad G_{\ell N} \Heq \frac{Q(v)}{R_H^4} \label{GLNH}
\end{align}
for some functions $R_H(v)$ (dimensions of length) and $Q_H(v)$ (dimensions of length squared) where the form of the latter is chosen for future convenience.
Then 
\begin{equation}
C \Heq R_{H,v}
\end{equation}
and by (\ref{Rp})
\begin{equation}
G_{\ell \ell } \Heq R_{H,v} \left( \frac{1}{R_H^2} - \frac{Q}{R_H^4} \right) \label{GLLH}
\end{equation}
Finally by (\ref{kappaH}),
\begin{equation}
\kappa \Heq C_{\rho} \Heq \frac{1}{2R_H} \left(1 - \frac{Q}{R_H^2} \right)  \, . \label{kappaHRN}
\end{equation}

\subsection{Off the horizon}

Next, integrate away from $\bar{H}$. First with  $G_{\! N \! N}=0$ (\ref{fcevol1}) can be integrated with initial condition (\ref{RIC}) to give
\begin{equation}
R(v,\rho) = R_H(v) - \rho \, .
\end{equation}
Then with (\ref{GLNH}) we can integrate (\ref{constd1}) to find
\begin{align}
G_{\ell N} & = \frac{Q}{R^4}  
\end{align}
and use this result and (\ref{GLLH}) to integrate(\ref{constd2}) to get
\begin{align}
%{\cal L}_N (R^2 G_{\ell \ell}) & = \frac{2 Q Q_v}{R^2} \\
G_{\ell \ell} & = \frac{\left(R_H^2-Q \right) R_{H,v}}{R_H^2 R^2} + \frac{\rho \, Q_{,v} }{R_HR^3} \, . 
\end{align}
With these results in hand and initial condition $R_\ell \Heq 0$ we integrate (\ref{fcevol2}) to get
\begin{equation}
R_\ell = \frac{\rho(Q-R_H^2+ \rho R_H )}{2R^2 R_H }
\end{equation}
and finally with initial conditions (\ref{Rp}) and (\ref{kappaH}) we can integrate (\ref{fcevol3}) to find
\begin{equation}
C = R_{H,v} - R_\ell \, .  \label{TF2}
\end{equation}

\subsection{Comparison with Vaidya-Reissner-Nordstr\"om}

We can now compare this derivation to a known example. The Vaidya-Reissner-Nordstr\"om  (VRN)  metric takes the form
\begin{equation}
\mbox{d} s^2 = - \left(1 - \frac{2m(v)}{r}+\frac{q(v)^2}{r^2} \right) \mbox{d}v^2 + 2 \mbox{d} v \mbox{d} r + r^2 \mbox{d} \Omega^2
\end{equation}
where the apparent horizon $r_H = m + \sqrt{m^2 - q^2}$ and $r$ is an affine parameter of the ingoing null geodesics. To put it into the form 
of (\ref{metant}) where the affine parameter measures distance off the horizon we make the transformation
\begin{equation}
r = r_H - \rho
\end{equation}
whence the metric takes the form
\begin{align}
\mbox{d} s^2 =& - \left(2 r_{H,v} - \frac{\rho \left(q^2 - r_H (r_H-\rho) \right) }{r_H (r_H-\rho)^2} \right) \mbox{d}v^2  \\& - 2 \mbox{d} v \mbox{d} \rho + (r_H - \rho)^2 \mbox{d} \Omega^2 \, . \nonumber
\end{align}
That is 
\begin{align}
C & =  r_{H,v} - \frac{\rho \left(q^2 - r_H (r_H-\rho) \right) }{2r_H (r_H-\rho)^2}\\
R & = r_H - \rho 
\end{align}
and on the horizon
\begin{align}
C \Heq r_{H,v} \qquad \mbox{and} \qquad R  \Heq r_H 
\end{align}
as expected. 

To do a complete match we calculate the rest of the quantities. First appropriate null vectors are
\begin{align}
\ell &=   \frac{\partial}{\partial v}  + \left(r_{H,v} - \frac{\rho \left(q^2 - r_H (r_H-\rho) \right) }{2r_H (r_H-\rho)^2}\right) \frac{\partial}{\partial \rho}  \\  
N &= \frac{\partial}{\partial \rho} \, . 
\end{align} 
Then direct calculation shows that 
\begin{align}
R_\ell & = - \frac{\rho \left(q^2 - r_H (r_H-\rho) \right) }{2r_H (r_H-\rho)^2}  \\
R_N & =  - 1
\end{align}
and
\begin{align}
G_{\ell \ell} & =  \frac{\left(r_H^2-q^2 \right) r_{H,v}}{r_H^2 r^2} + \frac{2 \rho q  q_{,v} }{r_H r^3}  \\
G_{\ell N} & =  \frac{q^2}{(r_H-\rho)^2} \\
G_{NN} & = 0 \\
G_{q} & =  \frac{2q^2}{(r_H-\rho)^2}  \; . 
\end{align}
It is clear that with $R_H = r_H$ and $Q=q^2$ our general results (\ref{GLNH})-(\ref{TF2}) give rise to the VRN spacetime (as they should). 

As expected the data on the horizon is sufficient to determine the spacetime everywhere back out along the ingoing null geodesics: we simply solve a set of (coupled) ordinary differential equations along each curve. With the matter providing the only dynamics and that matter only moving inwards along the geodesics the problem is quite straightforward. 
In this case there is no need to specify extra data on $\mathcal{N}$. 

%In fact it is not hard to 

We now turn to the more interesting case where the dynamics are driven by a scalar field for which there will be both inward and outward fluxes of matter.

\section{Massless scalar field}
\label{sec:matter modelsII}

Spherical spacetimes containing a massless scalar field $\phi(v,\rho)$ are governed by the stress energy tensor 
given by, 
\begin{align}
T_{ab} & =  \nabla_a \phi \nabla_b \phi - \frac{1}{2} g_{ab} \nabla^c \phi \nabla_c \phi 
\label{stress_sc}
\end{align}
This system has nonvanishing inward and outward fluxes which are,
\begin{align}
T_{\ell \ell} &= (\phi_{\ell})^2 \\ 
 T_{NN} & = (\phi_{N})^2.
\end{align}
Here and in the following keep in mind that $N = \frac{\partial}{\partial \rho}$ and so $\phi_N = \phi_{, \rho}$. 
We also observe from \eqref{stress_sc} that,
\begin{equation}
T_{\ell N} = 0.
\end{equation}
These fluxes are related by the wave equation%\footnote{This may be independently posited or, given the stress energy tensor,  recovered from the Bianchi identities.}:
\begin{equation}
\Box_g \phi := \nabla^\alpha \nabla_\alpha \phi = 0 \implies 
 { \left(R \phi_\ell \right)_{,\rho} = - R_\ell \phi_{,\rho } }. 
\label{wave_eq}
\end{equation}  
For our purposes we are not particularly interested in the value of $\phi$ itself but rather in the associated net flux of energies in the ingoing and outgoing null direction.
Hence we define
\begin{align}
\Phi_\ell = \sqrt{4 \pi} R \phi_{\ell} \quad \mbox{and} \quad \Phi_N = \sqrt{4 \pi} R \phi_{N}  \, .
\end{align}
Respectively these are the square roots of the scalar field energy fluxes in the $N$ and $\ell$ directions. 
That is, over a sphere of radius $R$, $\Phi_\ell$ is the square root of the total integrated flux in the $N$-direction and $\Phi_N$ is the square root of the total integrated flux
in the $\ell$-direction. Though not strictly correct, we will often refer to $\Phi_\ell$ and $\Phi_N$ themselves as fluxes.

Then (\ref{wave_eq}) becomes
\begin{align}
\Phi_{\ell, \rho} = - \frac{R_\ell \Phi_N}{R} \,  \label{wave_eqII}
\end{align}
or, making use of the fact that $\phi_{,v \rho} = \phi_{,\rho v}$, 
\begin{align}
\Phi_{N,v} = - \kappa \Phi_N - C \Phi_{N,\rho} - \frac{R_N \Phi_\ell}{R} \label{intCon} \, . 
\end{align}
These can usefully be  understood as  advection equations with sources. Recall that a general homogeneous advection equation can be written in the form
\begin{align}
\frac{\partial \psi}{\partial t} + C \frac{\partial \psi}{\partial x} = 0
\end{align}
where $C$ is the speed of flow of $\psi$: if $C$ is constant then this has the exact solution
\begin{equation}
\psi = \psi (x-Ct) 
\end{equation}
and so any pulse moves with speed $\frac{dx}{dt} = C$. Any non-homogeneous term  corresponds to a source which adds or removes energy from the system.  Then 
(\ref{wave_eqII}) tells us that the flux in the $N$-direction ($\Phi_\ell)$ is naturally undiminished as it flows along a (null) surface of constant $v$ and increasing $
\rho$. However the interaction with the flux in the $\ell$ direction can cause it to increase or decrease. Similarly (\ref{intCon}) describes the flow of the flux in the 
$\ell$-direction ($\Phi_N$) along a surface of constant $\rho$ and increasing $v$. Rewriting with respect to the affine derivative
(see Appendix \ref{daff})
%\footnote{\label{affineDerivative} This is equivalent to taking the derivative  with respect to a rescaling of $v$ that would set $\kappa = 0$. However the connection term varies with the
%quantity being differentiated. In contrast to the derivative for $\Phi_N$, for a quantity such as $\Phi_\ell$  that is linearly %dependent on $\ell$  it is 
% $D_v \Phi_\ell = \partial_v \Phi_\ell - \kappa \Phi_\ell$. For more details see \cite{Booth:2006bn}. } 
$D_v = \partial_v + \kappa$
 it becomes
\begin{align}
D_v \Phi_N + C \Phi_{N,\rho} = - \frac{R_N \Phi_\ell}{R} \label{adN} \, . 
\end{align}
Then, as might be expected, $\Phi_N$ naturally flows with coordinate speed $C$ (recall that $\ell = \frac{\partial}{\partial v} + C \frac{\partial}{\partial \rho}$ so this is the speed of 
outgoing light relative to the coordinate system) but its strength can be augmented or diminished by interactions with the outward flux.

\subsection{System of first order PDEs}
\label{firstsystem}
Together (\ref{wave_eqII}) and (\ref{intCon}) constitute a first-order system of partial differential equations  for the scalar field. We now restructure the gravitational field equations in the same way. 

First with respect to $\Phi_\ell$ and $\Phi_N$ the  constraint equations (\ref{const1})-(\ref{const3}) on constant $\rho$ surfaces become:
\begin{align}
R_{,v} &= R_\ell - C R_N \label{consc1} \\
R_{\ell,v} &= \kappa R_{\ell} +  \frac{C \left( 1 + 2 R_{\ell} R_N \right)}{2 R} - \frac{\Phi_\ell^2}{R} \label{consc2} \\   
R_{N,v} &= -\kappa R_{N} -  \frac{\left( 1 + 2  R_{\ell} R_N \right)}{2 R} +   \frac{C \Phi_N^2}{R} \label{consc3} %\phi_{N,v} &= - (C \phi_{N})_{,\rho} -  \frac{\left(  R_\ell \phi_{N} + R_N \phi_{\ell} \right)}{ R} \label{consc4}
\end{align}
while the ``time''-evolution equations (\ref{cevol1})-(\ref{cevol3}) are:
\begin{align}
& {R_{,\rho\rho}  = - \frac{\Phi_N^2}{R}}  \label{C1} \\
&  { \left( R R_{\ell}\right)_{,\rho} = - \frac{1}{2} }\label{C2a} \\
&  {C_{,\rho \rho}}\,\, {= \frac{1 + 2 R_\ell R_N}{R^2} - \frac{2  \Phi_\ell \Phi_N}{R^2} }  \, . \label{C3}
\end{align}

Two of these equations can be simplified. First, integrating  (\ref{C2a}) from $\rho = 0$ on which 
$R_\ell \Heq 0$ we find
\begin{align}
R_\ell = - \frac{ \rho}{2 R} \label{Rl} \, . 
\end{align}
This can be substituted into (\ref{consc2}) to turn it into an algebraic constraint
\begin{align}
C = 2  \Phi_\ell^2 - 2 R_\ell\left(\kappa R + R_\ell \right) \label{C2}  \, .%+ 2 (R^o R^o_\ell)_{,v}  \; .  
%\label{CEq}
\end{align}
%The last term will vanish for a MOTT. 

Despite these simplifications, the presence of interacting  outward and inward matter fluxes means that in contrast to the dust examples,
this is truly a set of coupled partial differential equations. 
Hence we can expect that the matter and spacetime  dynamics will be governed by off-horizon data in addition to data at $\rho = 0$.

We  reformulate as a system of first order PDEs in the following way. 
First designate 
\begin{align}
  \; \; \{R, R_N, \kappa, \Phi_\ell, \Phi_N \} 
\end{align}
as the \emph{primary variables}. The \emph{secondary variables} $\{R_\ell, C\}$ are defined by (\ref{Rl}) and (\ref{C2}) in terms of the primaries. 
%\begin{align}
%R_\ell  & = -\frac{ %2 R^o R^o_\ell - 
%\rho}{2 R}  \; \;   \mbox{and} \label{RLEq} \\
%C & = 2 \Phi_\ell^2 - 2 R_\ell\left(\kappa R + R_\ell \right) %+ 2 (R^o R^o_\ell)_{,v} 
%
% \end{align}
%are defined in terms of the primaries. 

Next on $\rho=\mbox{constant}$ surfaces the primary variables are constrained by 
\begin{align}
R_{,v} &= R_\ell - C R_{N}  \; \;  \mbox{and} \label{REq} \\
R_{N, v} &= -\kappa R_N -  \frac{ 1}{2 R} \left( 1 + 2  R_{\ell} R_{N} - 2 C \Phi_N^2\right) \label{RNVEq}
\end{align}
along with scalar flux equation (\ref{intCon}) while their time evolution is governed by 
\begin{align}
R_{,\rho} &= R_N \label{RNEq} \\
R_{N,\rho} & =  - \frac{\Phi_N ^2}{R} \label{RNNEq} \\
\kappa_{,\rho} & =   \frac{1}{R^2} \left(1 + 2 R_\ell R_{N} - 2 \Phi_{\ell} \Phi_{N}\right)  \label{kappaEq}   \\
 { \Phi}_{\ell,\rho} &= -  \frac{R_\ell \Phi_N}{R} \label{PhiLrhoEq} \, .  \, 
\end{align}
%As for (\ref{adN}), the last equation (\ref{PhiLrhoEq}) can be understood as a non-homogeneous advection equation, though the flow is in the $N = \partial/\partial \rho$ direction (with 
%zero speed relative to the coordinates). The two are coupled with the right-hand sides governing energy exchanges between the fluxes. 
%
%On a practical level, the reader will notice that we are missing a time evolution equation for $\Phi_N$. The integration condition (\ref{intCon}) could be rearranged into that form but 
%at the price of a division by $C$: in which case that evolution equation is ill-defined for isolated horizons. This is not so surprising as we do not expect
%a well-defined initial value formulation based on data specified on a single null
%surface. We certainly wish to allow for isolated horizons as possible configuration and so propose the following integration scheme\footnote{For $C \neq 0$ and so $\bar{H}$ spacelike we would expect a good initial value formulation of the usual $(3+1)$-type. However, as discussed in the introduction, the region of spacetime determined is necessarily restricted to lie within the event horizon. Hence even in that case it is desirable to include extra data from $\mathcal{N}$.  }.   
%Winicour:2013gha}). {\bf -- this may not be true anymore?}

We now consider how all of these equations may be used to integrate final data. The scheme is closely related to that used in \cite{Winicour:2012znc}.

\subsection{Final data on $\bar{H}$ and $\bar{\mathcal{N}}$} 
\label{flux:onh}

In line with the depiction in FIG.\ref{hd2}, 
we specify  final data on $H \cup {\cal N}$ or rather on the sections $\bar{H} \cup \bar{{\cal N}}$ where
\begin{align}
\bar{H} &= \{(0,v)\in H: v_i \leq v \leq v_f\} \; \; \mbox{and} \\
\bar{\mathcal{N}} & = \{(\rho,v_f) \in \mathcal{N}: \rho_i \leq \rho \leq 0 \} \nonumber \, . 
\end{align}
Their intersection sphere is $\bar{H} \cap \bar{{\cal N}} = (0,v_f)$. Here and in what follows we suppress the angular coordinates. 

 The final data is
\begin{align}
\bar{H}:&\; \; \Phi_\ell  \label{data} \\
\bar{{\cal N}}:&\;\; \Phi_N  \nonumber  \; \; \mbox{and} \\
\bar{H}  \cap \bar{\cal N}:& \;\; R=R_o \nonumber \; . 
\end{align}
$\Phi_\ell$ on $\bar{H}$ is a function of $v$ while  $\Phi_N$ on $\mathcal{\bar N}$ is a function of $\rho$. $R_o$   is a single number. 

Further on ${H}$ we have
\begin{align}
R_\ell \Heq 0 \; \; \mbox{and} \; \;  R_N \Heq -1
\end{align}
where the null vectors are scaled in the usual way and, as before,   the 
notation  $\Heq$  indicates that all quantities on both sides of the 
equality are evaluated on $H$. 

This data can be used to evaluate $C$ and $R$ on $\bar{H}$. From (\ref{C2}) and  (\ref{REq})
\begin{align}
C \Heq & \,  2 \Phi_\ell^2  \label{Feq1} \; \; \mbox{and}\\ 
R \Heq &  \, R_o + 2\int_{v_f}^{v} \! \!  \Phi_\ell^2 \, \mbox{d} v  \, . \label{Rdyn1}
\end{align}

To find $\Phi_N$ on $\bar{H}$ we would need to solve
\begin{align}
\Phi_{N,v} +  \frac{1}{2R} \left( 1 - 4 \Phi_\ell^2 \Phi_N^2 \right) \Phi_N \Heq - 2 \Phi_\ell^2 \Phi_{N,\rho} + \frac{\Phi_\ell}{R} \label{PhiNFV}
\end{align}
which comes from (\ref{intCon}) combined with the above results. However at this stage $\Phi_{N,\rho}$ isn't known and so this can only be solved
directly in the isolated $\Phi_\ell \Heq 0$ case. There
\begin{align}
\Phi_N^{\mbox{\tiny{iso}}} \Heq \Phi_{N_f} e^{- \nicefrac{(v - v_f)}{2R_o}} \label{Phi_solved}
\end{align}
where $\Phi_{N_f} = \Phi_N(0, v_f)$. Equivalently (see Appendix \ref{daff})  $\Phi_N$  is affinely constant on an isolated horizon. 

With $R_N=-1$,  (\ref{consc3}) tells us that 
\begin{align}
\kappa \Heq & \, \frac{1}{2R} \left( 1 - 2C \Phi_{N}^2 \right)   \, ,  \label{Feq3}
\end{align}
and so without $\Phi_N$ on $\bar{H}$ we also can't determine this (away from isolation). However the corner  $\bar{H}  \cap \bar{\cal N}$  is an 
exception to that rule. There we know $\Phi_\ell$, $\Phi_N$ and $R_o$ and so
\begin{align}
\kappa \eqq{\bar{H}  \cap \bar{\cal N}}    \, \frac{1}{2R_o} \left( 1 - 4 \Phi_\ell^2 \Phi_{N}^2 \right)  \, . \label{kappaHN}
\end{align}

The situation is less complicated on $\bar{\mathcal{N}}$. There with $\Phi_N$ as known data and final values known for all quantities on 
$\bar{H} \cap \bar{\mathcal{N}}$ all other quantities can be calculated in order
\begin{enumerate}[i)]
\item  Solve (\ref{RNEq}) and (\ref{RNNEq}) for $R$ and $R_N$.  \label{i}
\item  Calculate $R_\ell$ from (\ref{Rl}).
\item  Solve (\ref{PhiLrhoEq}) for $\Phi_\ell$.
\item  Solve (\ref{kappaEq}) for $\kappa$. 
\item  Calculate  $C$ from (\ref{C2}).  \label{vi}
\end{enumerate}
We then have all data on $\bar{\cal N}$.
\begin{figure}[h]
\begin{center}
\includegraphics[scale=0.3]{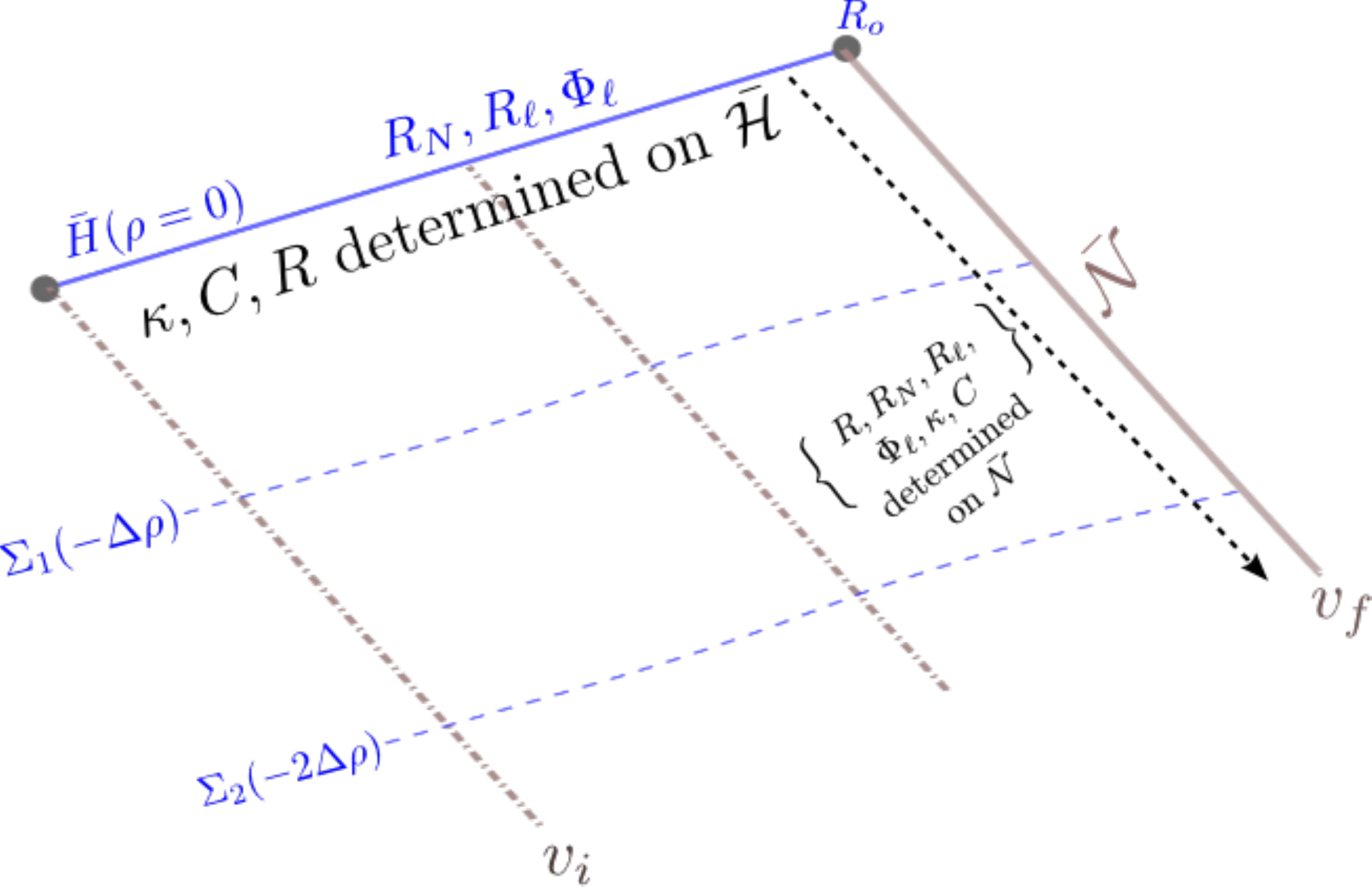} 
\caption{ The constraint equations along with initial conditions on the horizon, i.e. $R_\ell \Heq 0, R_N \Heq -1 $ determine $\kappa$, $C$ and $R$ on $\bar{H}$}
\label{fig:onh}
\end{center}
\end{figure}

\subsection{Integrating from the final data}
\label{sec:int}

We now consider how that data can be integrated into the causal past of $\bar{H} \cup \bar{\mathcal{N}}$. The basic steps in the integration 
scheme are
%\begin{enumerate}[a)]
%\item integrating $\Phi_N$ along surfaces of constant $\rho$ using (\ref{intCon}), 
%\item calculating $\Phi_{N,\rho}$ by comparing with a $\rho$ surface on which $\Phi_N$ is already known
%\item integrating $\{R, R_N \Phi_L, \kappa\}$ along surfaces of constant $v$ using (\ref{REq})-(\ref{PhiLrhoEq}) and 
%\end{enumerate} 
%However the equations are all coupled and so the integrations interlock. In particular calculating of $\Phi_{N,\rho}$ and its appearance
%in $\Phi_{N,v}$ stymies many analytic investigations of these solutions.
%
%The steps are 
demonstrated in a simple numerical integration based on Euler approximations. 
This scheme alternates between using steps \ref{i})-\ref{vi}) to integrate data down the characteristics of constant $v$ followed by an application of (\ref{intCon}) to calculate $\Phi_N$ on the next characteristic. 

\begin{figure}[t]
\begin{center}
\includegraphics[scale=0.25]{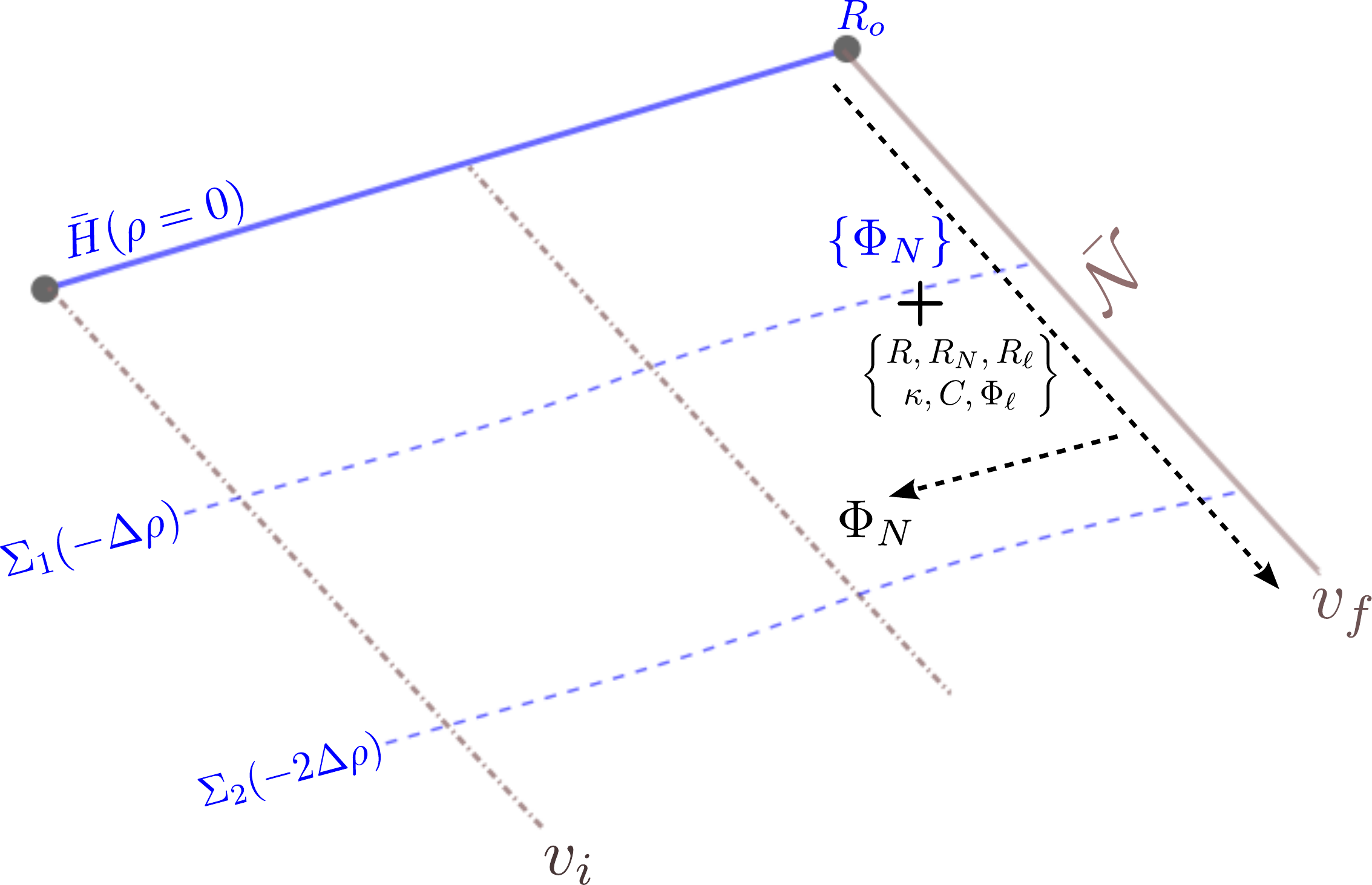} 
\caption{Evolving $\Phi$ in the $- \frac{\partial}{\partial v}$  direction. }
\label{fig:onh1}
\end{center}
\end{figure}

In more detail, assume a discretization $\{v_m, \rho_n\}$ (with $m$ and $n$ at their maxima along the 
final surfaces) by steps $\Delta v$ and $\Delta \rho$. Then if all data is known along a surface  $v_{m+1}$ and $R$ and $\Phi_\ell$ are known
everywhere on  $\bar{H}$:
%
%
%
%Then if all data is known at $(v_m, \rho_{n+1})$ and $(R, R_N, R_\ell, \kappa, \Phi_\ell)$ are known at 
%$(v_{m+1}, \rho_n)$ we can find all data at $(v_m, \rho_n)$. This is done by the following procedure (illustrated  in FIG.~\ref{vmrhon}).
%\begin{figure}
%\includegraphics[scale=0.6]{vmrhon}
%\caption{How data is calculated at $(v_m, \rho_n)$: a) $\Phi_N$ is approximated from data at $(v_{m+1}, \rho_n)$  b)  $\Phi_{N,\rho}$ is approximated by comparing
%$\Phi_N$ at $(v_m, \rho_n)$ and $(v_{m}, \rho_{n+1})$ and  c)  $R$, $R_N$, $\Phi_\ell$ and $\kappa$ are approximated from data at 
%$(v_m, \rho_{n+1})$.}
%\label{vmrhon}
%\end{figure}
\begin{enumerate}[a)]
\item Use the knowledge of $\Phi_N$ on $v_{m+1}$ to calculate $\Phi_{N,\rho}$. 
\item Use (\ref{intCon}) at $(v_{m}, \rho_n)$ to find $\Phi_{N,v}$. Then
\begin{align}
\Phi_N(v_{m}, \rho_n) \approx \Phi_N(v_{m+1}, \rho_n) -  \Phi_{N,v}(v_{m+1}, \rho_n)  \Delta v
\end{align}
%\item Approximate 
%\begin{align}
%\Phi_{N,\rho} (v_{m}, \rho_n) \approx \frac{\Phi_{N} (v_{m}, \rho_{n+1}) - \Phi_{N} (v_{m}, \rho_n)}{\Delta \rho}
%\end{align}
\item Apply  %this new knowledge of $\Phi_N$ in  
(\ref{Feq3}) to calculate $\kappa$ at $(v_{m}, 0)$. 
\item Use (\ref{RNEq})-(\ref{PhiLrhoEq}) to integrate the values of $R_{N,\rho}$,  $\kappa_{,\rho}$ and $\Phi_{\ell,\rho}$ out along the 
$v = v_m$ characteristic as for the initial data. 
%
%
% at $(v_m, \rho_{n+1})$ to find $R_{,\rho}$, $R_{N,\rho}$,  $\kappa_{,\rho}$ and $\Phi_{\ell,\rho}$. Then
%generically if $X$ is any of these quantities
%\begin{align}
%X(v_{m}, \rho_n) &\approx X(v_{m}, \rho_{n+1}) -  X_{,\rho}(v_{m}, \rho_{n+1})  \Delta \rho  
%\end{align}
\end{enumerate}
This can then be repeated marching all the way along $\bar{H}$ as shown in FIG.\ref{fig:onh1}.

This is how we would proceed for general cases. However those general studies will  be left for a future paper. Here instead we will focus on spacetime near a slowly evolving horizon. There, as will be seen in the next section, $\Phi_{N,\rho}$ is negligible and it becomes possible to integrate along surfaces of constant $v$. 

%
%more useful to integrate backwards along the surfaces of constant $\rho$ rather than out along the surfaces of constant $v$ (as in FIG.~\ref{fig:onh} but on other $\rho$ surfaces). In this case we repeatedly apply steps a)-c) to 
%integrate in the $- \frac{\partial}{\partial v}$ direction, stopping at each step to calculate $\Phi_{N,\rho}$ to use in the evolution of $\Phi_N$ to the next $v_m$. As we shall see in the next section, in the slowly evolving, near horizon case things simplify: these pauses are not necessary as the $\Phi_{N,\rho}$ term will become negligible. Hence one can integrate continuously in the $v$-direction. 
%
It may not be  immediately obvious how this integration scheme obeys causality and what restricts it 
to determining points inside the domain of dependence. 
This is  briefly discussed in Appendix \ref{causalityApp}. %demonstrates how causality plays into this integration scheme.

%
%
%Start with the fully integrated final data on $\bar{H}$ and $\bar{\mathcal{N}}$ and assume a discretization $\{v_m, \rho_n\}$ (with $m$ and $n$ 
%at their maxima along the final surfaces) by steps 
%$\Delta v$ and $\Delta \rho$. We begin at $(v_m, \rho_n) = ( v_f- \Delta v, -\Delta \rho)$. Then with known data at $(v_m, \rho_{n+1})$ and 
%$(v_{m+1}, \rho_n)$ we can:
%\begin{enumerate}[a)]
%\item Use (\ref{intCon}) at $(v_{m+1}, \rho_n)$ to evolve 
%\begin{align}
%\Phi_N(v_{m+1}, \rho_n) \rightarrow \Phi_N(v_{m}, \rho_n)
%\end{align}
%\item Approximate 
%\begin{align}
%\Phi_{N,\rho} (v_{m}, \rho_n) \approx \frac{\Phi_{N} (v_{m}, \rho_{n+1}) - \Phi_{N,\rho} (v_{m}, \rho_n)}{\Delta \rho}
%\end{align}
%\item Use (\ref{REq})-(\ref{PhiLrhoEq}) at $(v_m, \rho_{n+1})$ to evolve all other quantities from there to $(v_m, \rho_{n})$
%\end{enumerate}
%We then know all data at $(v_m, \rho_n)$ and so can move on to either $(v_m, \rho_{n-1})$ or $(v_{m-1}, \rho_n)$
%
\subsection{Spacetime near a slowly evolving horizon}
We now apply the formalism to a concrete example: weak scalar fields near the horizon. Physically
the black hole will be close to equilibrium and hence the horizon slowly evolving in the sense of  \cite{Booth:2003ji,Booth:2006bn}.

``Near horizon'' means
 that we expand all quantities as Taylor series in $\rho$ and keep terms up to order $\rho^2$. ``Weak scalar field'' 
means that we assume 
\begin{align}
\Phi_N, \Phi_{\ell} \sim  \frac{\varepsilon}{R}
\end{align}
and then expand the terms of the Taylor series up to order $\epsilon^2$. To order $\epsilon^0$ 
the spacetime will be vacuum (and Schwarzschild), order $\epsilon^1$ will be a test scalar field propagating on 
the Schwarzschild background and order $\epsilon^2$ will include the back reaction of the scalar field on the 
geometry. 

\subsubsection{Expanding the equations}
We expand all quantities as Taylor series in $\rho$. That is for $X \in \{ R, R_N, R_\ell, \kappa, C, \Phi_\ell, \Phi_N\}$
\begin{align}
X(v, \rho) = \sum_{n = 0}^{\infty} \frac{ \rho^nX^{(n)}(v)}{n!} 
\end{align}
with 
\begin{align}
R_N^{(n)} = R^{(n+1)} \; \; \mbox{and} \; \kappa^{(n)} = C^{(n+1)} \; . 
\end{align}

%\begin{align}
%R  &= \sum_{n = 0}^{\infty} \frac{ \rho^nR^{(n)}}{n!}\\
%R_N  &= \sum_{n = 0}^{\infty} \frac{ \rho^nR_N^{(n)}}{n!} \\
%R_\ell  &= \sum_{n = 0}^{\infty} \frac{ \rho^nR_\ell^{(n)}}{n!} \\
%\kappa  &= \sum_{n = 0}^{\infty} \frac{ \rho^n\kappa^{(n)}}{n!}   \\
%C  & = \sum_{n = 0}^{\infty} \frac{ \rho^n C^{(n)}}{n!}   \\
%\Phi_\ell  &=  \sum_{n = 0}^{\infty} \frac{\rho^n \Phi^{(n)}_{\ell}(v)}{n!} \\
%\Phi_N &=  \sum_{n = 0}^{\infty} \frac{\rho^n \Phi^{(n)}_{N}(v)}{n!}  \, . 
%\end{align}
The free final data is $\Phi_\ell^{(0)}$ on $\bar{H}$, $R_o$ on $\bar{H} \cap \bar{\mathcal{N}}$ and the 
Taylor expanded
\begin{align}
\Phi_{N_f} (\rho) = \sum_{n=0}^{\infty} \frac{\rho^n}{n!} \Phi_{N_f}^{(n)} \,  \label{PNf}
\end{align}
on $\bar{\mathcal{N}}$.
%The final data are $R_o$,  $\Phi_{\ell_f}^{(0)}$ and the  $\Phi_{N_f}^{(n)}$ from (\ref{PNf}). 
Following  \cite{Frolov:1998wf} we give names to special cases of this free data:
\begin{enumerate}[i)]
\item \emph{out-modes}: no flux through $\bar{H}$ ($ \Phi_{\ell}^{(0)} = 0$),\\
non-zero flux through $\bar{\mathcal{N}}$ ($\Phi^{(n)}_{N} \neq 0$ for some $n$)
\item \emph{down-modes}: non-zero flux through $\bar{H}$ ($\Phi_{\ell}^{(0)} \neq 0$), \\
zero flux through ${\bar{\mathcal{N}}}$ ($\Phi^{(n)}_{N} = 0$ for all $n$)
\end{enumerate}

From the free data we construct the rest of the final data on $\bar{H}$. Equations (\ref{Feq1}) and (\ref{Feq3}) give
\begin{align}
C^{(0)} & = 2 \Phi_\ell^{(0)^{\mbox{\scriptsize{2}}}} \\
C^{(1)} & = \kappa^{(0)} \approx \frac{1}{2 R^{(0)}} \, . 
\end{align}
Here and in what follows the $\approx$ indicates that terms of order $\epsilon^3$ 
or higher have been dropped. Further by our gauge choice
\begin{align}
R_N^{(0)} = R^{(1)} = -1 
\end{align}
and so from (\ref{Rdyn})
\begin{align}
R^{(0)} =  R_o + \int_{v_f}^{v} \! \!  C^{(0)} \, \mbox{d} v  \, . \label{Rdyn}
\end{align}
This is an order $\epsilon^2$ correction as long as the interval of integration is small relative to 
$\nicefrac{1}{\epsilon}$.

The last piece of final data on $\bar{H}$  is $\Phi_N^{(0)}$ and comes from the first order differential equation (\ref{PhiNFV}) 
\begin{align}
\frac{\mbox{d}\Phi_N^{(0)}}{\mbox{d} v}  + \frac{\Phi_N^{(0)}}{2R_o}  \approx \frac{ \Phi_\ell^{(0)} }{R_o} \label{PN1} \, 
\end{align}
which has the solution
\begin{align}
\Phi_N^{(0)} = \Phi_{N_f}^{(0)}e^{\nicefrac{(v_f - v)}{2R_o}} + e^{-\nicefrac{v}{2R_o}} \int_{v_f}^v e^{\nicefrac{\tilde{v}}{2R_o}} \Phi_\ell^{(0)} \mbox{d} \tilde{v} \label{PLInt}
\end{align}
in which the free data $\Phi_{N_f}^{(0)}$ came in as a boundary condition. 
Note that scalar fields that start small on the boundaries remain small in the interior, again as long as 
the integration time is short compared to $\nicefrac{1}{\epsilon}$. We assume that this is the case. 

From the final data, the black hole is close to equilibrium and the horizon is slowly evolving to order
 $\epsilon^2$. That is, the expansion parameter \cite{Booth:2003ji,Booth:2006bn}:
\begin{align}
C \left( \frac{1}{2}  \theta_{(N)}^2 + G_{ab} N^a N^b \right) \approx  \left(\frac{4 \Phi_\ell^2 }{R^2} \right)  \sim \frac{4\epsilon^2}{R^2}  \; . 
\end{align}
Further we already have the first order expansion of $C$:
\begin{align}
C \approx  2 \Phi_\ell^{(0)^{\mbox{\scriptsize{2}}}} \!  + \frac{\rho}{2 R_o} \, . 
\end{align}
That is (to first order) there is a null surface at
\begin{align}
\rho_{\mbox{\tiny{EHC}}} \approx  - 4 R_o  \Phi_\ell^{(0)^{\mbox{\scriptsize{2}}}} \; . 
\end{align}
This null surface  is the event horizon candidate discussed in \cite{Booth:2012xm}: if the horizon remains slowly evolving throughout its future evolution and ultimately transitions to isolation then 
the event horizon candidate is the  event horizon. 

Moving off the horizon to calculate up to second order in $\rho^2$, from (\ref{RNEq}) and (\ref{RNNEq}) we find
\begin{align}
%R^{(0)} & = R_o \\
%R_N^{(0)}  =  R^{(1)} & = -1 \\
R_N^{(1)}  = R^{(2)} & \approx  - \frac{\Phi_{N}^{{(0)}^{\mbox{\scriptsize{2}}}}}{R_o} \label{RX2} \\
R_N^{(2)}  & \approx  -\frac{\Phi_N^{(0)} \left(\Phi_N^{(0)}+2R_o \Phi_N^{(1)} \right)}{R_o^2} \,
\label{RX3}  
\end{align}
and so from (\ref{Rl}) 
\begin{align}
R_\ell^{(0)} & = 0 \\
R_\ell^{(1)} & = - \frac{1}{2R^{(0)}} \\
R_\ell^{(2)} & = -  \frac{1}{R^{{(0)}^{\mbox{\scriptsize2}}}} \, . 
\end{align}
Note that the last two terms will include terms of order $\epsilon^2$ once the (\ref{Rdyn}) integration is done to calculate 
$R^{(0)}$. 

From (\ref{PhiLrhoEq}) we can rewrite $\Phi_\ell^{(n)}$ terms with respect to $\Phi_N^{(n)}$ ones:
\begin{align}
\Phi_\ell^{(1)} &= 0 \label{PL1} \\
\Phi_\ell^{(2)} & \approx \frac{ \Phi^{(0)}_N}{2 R_o^2} \label{PL2} \, .
%\Phi_\ell^{(3)} &= \frac{2 \Phi_N^{(0)} + \Phi_N^{(1)} R_o }{R_o^3} \, .
%\Phi_\ell^{(4)} &= \frac{3 \left(6 \Phi_N^{(0)} + 4 R_o \Phi_N^{(1)} + R_o^2 \Phi_N^{(2)}  \right) }{2R_o^4}
\end{align}
The vanishing linear-order term reflects the fact that close to the horizon (where $R_\ell = 0$) the inward 
flux decouples from the outward (\ref{PhiLrhoEq}) and so freely propagates into the black hole. Physically this means that (to first 
order in $\rho$ near the horizon) the horizon flux is approximately equal to the ``near-horizon'' flux. 

Next, from  (\ref{kappaEq}) 
%\begin{align}
%\kappa^{(1)} & \approx\frac{1}{R_o^2}\left( 1 - \Phi_\ell^{(0)} \Phi_N^{(0)}  \right) \mbox{  and} \\
%\kappa^{(2)} & \approx  \frac{1}{R_o^3} \left( 3 - 2 \Phi_\ell^{(0)} \left(2 \Phi_N^{(0)} + R_o    \Phi_N^{(1)}\right) \right)  \, ,
%%\kappa^{(3)} & = \frac{2}{R^{{(0)}^{\mbox{\scriptsize{4}}}}}\Bigg(6   -  6 \Phi_\ell^{(0)} \Phi_N^{(0)} - 2 R^{{(0)}}  \left( R^{(2)} + 2 \Phi_\ell^{(0)} \Phi_N^{(1)} \right) \nonumber\\
%%& \phantom{xxxxxxxx}-   R^{{(0)}^{\mbox{\scriptsize{2}}}} \left(\Phi_N^{(2)} \Phi_\ell^{(0)}+\Phi_N^{(0)}\Phi_\ell^{(2)} \right)  \Bigg) 
%\end{align}
%
\begin{align}
\kappa^{(1)} & =  C^{(2)}  \approx\frac{1}{R^{{(0)}^{\mbox{\scriptsize2}}}} -\frac{ 2 \Phi_\ell^{(0)} \Phi_N^{(0)}}{R_o^2}   \mbox{  and} \\
\kappa^{(2)}  & \approx  \frac{3}{R^{{(0)}^{\mbox{\scriptsize2}}}} - \frac{2\Phi_\ell^{(0)} \left(2 \Phi_N^{(0)} + R_o    \Phi_N^{(1)}\right) }{R_o^2}   \, .
%\kappa^{(3)} & = \frac{2}{R^{{(0)}^{\mbox{\scriptsize{4}}}}}\Bigg(6   -  6 \Phi_\ell^{(0)} \Phi_N^{(0)} - 2 R^{{(0)}}  \left( R^{(2)} + 2 \Phi_\ell^{(0)} \Phi_N^{(1)} \right) \nonumber\\
%& \phantom{xxxxxxxx}-   R^{{(0)}^{\mbox{\scriptsize{2}}}} \left(\Phi_N^{(2)} \Phi_\ell^{(0)}+\Phi_N^{(0)}\Phi_\ell^{(2)} \right)  \Bigg) 
\end{align}
%and from (\ref{C2}) 
%\begin{align}
%C^{(1)} & = \frac{1}{2 R^{(0)}} \\
%C^{(2)} & = \frac{1}{2 R^{(0)^{\mbox{\scriptsize2} }}}  - \frac{\Phi_\ell^{(0)} \Phi_N^{(0)} }{R_o^2} 
%\end{align}
Again keep in mind that the $R^{(0)}$ terms will be corrected to order $\epsilon^2$ from (\ref{Rdyn}).

Finally these quantities may be substituted into (\ref{intCon}) to get differential equations for the 
$\Phi^{(n)}_N$:
\begin{align}
%\frac{\mbox{d}\Phi_N^{(0)}}{\mbox{d} v}  + \frac{\Phi_N^{(0)}}{2R_o}  &= \frac{ \Phi_L^{(0)} }{R_o} \label{PN1} \\
\frac{\mbox{d}  \Phi_N^{(1)} }{\mbox{d} v} + \frac{\Phi_N^{(1)}}{R_o}  & \approx \frac{ \Phi_{\ell}^{(0)}}{R_o^2} - \frac{\Phi_N^{(0)}  }{R_o^2}  \label{PN2} \\
\frac{\mbox{d} \Phi_N^{(2)}}{\mbox{d} v}  + \frac{3 \Phi_N^{(2)}}{2R_o}  & \approx \frac{2 \Phi_{\ell}^{(0)}}{R_o^3} - \frac{5 \Phi_N^{(0)}}{2 R_o^3} - \frac{3 \Phi_N^{(1)}  }{R_o^2}  \label{PN3}  \, . 
%\frac{\mbox{d} }{\mbox{d} v}  \Phi_N^{(3)}+ \frac{2 \Phi_N^{(3)}}{R_o}  &= \frac{12 \Phi_L^{(0)} - 17 \Phi_N^{(0)} -  22 R_o \Phi_N^{(1)}  - 12 R_o^2 \Phi_N^{(2)}  }{2R_o^4} \\
%\frac{\mbox{d} \Phi_N^{(3)}}{\mbox{d} v}  + \frac{2 \Phi_N^{(3)}}{R_o}  &= \frac{6 \Phi_L^{(0)}}{R_o^4} - \frac{17 \Phi_N^{(0)}}{2R_o^4} -  \frac{11 \Phi_N^{(1)}}{R_o^3}  - \frac{6 \Phi_N^{(2)}}{R_o^2}    \, . 
\end{align}
Like (\ref{PLInt}) these are easily solved with an integrating factor and respectively have $\Phi_{N_f}^{(1)}$ and $\Phi_{N_f}^{(2)}$ as boundary conditions. 

Note the important simplification in this regime that enables these straightforward solutions.  The fact that $R_\ell \sim \rho $ has raised the 
$\rho$-order of the $\Phi_{N,\rho}$ terms. As a result  we can integrate directly across the
$\rho=\mbox{constant}$ surfaces rather than having to pause at each step to first calculate the 
$\rho$-derivative. 
The $\Phi_{N_f}^{(n)}$ are final data for these equations. They can be solved order-by-order and then 
 substituted back into the other expressions to reconstruct the near-horizon spacetime.

It is also important that the matter and geometry equations decompose cleanly in orders of $\epsilon$: 
we can solve the matter equations at order $\epsilon$ relative to a fixed background geometry and 
then use those results to solve for the corrections to the geometry at order $\epsilon^2$. 

\subsubsection{Constant inward flux}

We now consider the concrete example of an affinely constant flux through $\bar{H}$ along with an analytic flux through 
$\bar{\mathcal{N}}$. 
Then by Appendix \ref{daff}
\begin{align}
\Phi_\ell^{(0)} = \Phi_{\ell_f}^{(0)}  {e}^{V} \, ,
\end{align}
where $ \Phi_{\ell_f}^{(0)} $ is the value of $\Phi_\ell^{(0)}$ at $v_f$ and $V = \frac{v-v_f}{2 R_o}$ while $\Phi_{N_f}$ retains
its form from (\ref{PNf}). 

%
%if $ \Phi_{\ell_f}^{(0)} = 0$ while at least some of the $\Phi^{(n)}_{N_f} \neq 0$ then we have an  \emph{out-mode} with a flux in the $\ell$ direction towards $\mathscr{I}^+$ but no
%flux through the horizon. Conversely if $ \Phi_{\ell_f}^{(0)} \neq 0$ while all of the $\Phi^{(n)}_{N_f} = 0$ then we have a \emph{down-mode} with the only boundary flux through the horizon. 
%Note that these are completely independent and do not constrain each other.

We  solve the equations for this data up to second order in $\rho$ and $\epsilon$. First for  $\Phi_N^{(n)}$ equations we find:
\begin{align}
\Phi_N^{(0)} \approx & \;     \left(e^V \!-e^{-V} \right)  \Phi_{\ell_f}^{(0)}   +   e^{-V} \Phi_{N_f}^{(0)}  \label{matter1} \\
\Phi_N^{(1)}  \approx &  \;  \frac{2 \Phi_{\ell_f}^{(0)} }{R_o} \left(1 - e^{-2V}  \right) + \frac{2\Phi_{N_f}^{(0)}}{R_o} \left(e^{-2V} - e^{-V} \right) \\
&  \! +  \Phi_{N_f}^{(1)} e^{-2V} \nonumber  \\
\Phi_N^{(2)} \approx & - \frac{\Phi_{\ell_f}^{(0)}}{4R_o^2}  \left(e^V \!+14 e^{-V} - 48 e^{-2V}\! + 33 e^{-3V}   \right)  \\
& +\frac{\Phi_{N_f}^{(0)}}{2R_o^2} \left(7e^{-V} \! - 24 e^{-2V} +17 e^{-3V} \right) \nonumber \\
& + \frac{6\Phi_{N_f}^{(1)}}{R_o} \left(e^{-3V} \!- e^{-2V}  \right) +  \Phi_{N_f}^{(2)} e^{-3V}
\end{align}
and so 
\begin{align}
\Phi_\ell^{(0)} & =  e^{V} \Phi_{\ell_f}^{(0)}\, \\
\Phi_\ell^{(1)} &= 0 \\
\Phi_\ell^{(2)} & \approx  \frac{ \Phi_{\ell_f}^{(0)} }{2R_o^2} \left(e^V \!-e^{-V} \right) +  \frac{ \Phi_{N_f}^{(0)}}{2R_o^2}  e^{-V}   \label{matter2}\, . 
\end{align}
The scalar field equations are linear and so it is not surprising that to this order in $\epsilon$
each solution can be thought of as a linear combination of down and out modes. 

However for the geometry at order $\epsilon^2$, down and out modes no longer combine in a linear way. These quantities
can be found simply by substituting the $\Phi_\ell^{(n)}$ and $\Phi_N^{(n)}$ into the expression for $R^{(n)}$, $R_N^{(n)}$, 
$R_\ell^{(n)}$, $C^{(n)}$ and $\kappa^{(n)}$ given in the last section. They are corrected at order $\epsilon^2$ by  flux  terms
that are quadratic in  combinations of $\Phi_{\ell_f}^{(m)}$ and $\Phi_{N_f}^{(n)}$. The  terms are somewhat messy 
and the details not especially enlightening. Hence we do not write them out explicitly here.

\subsubsection{$\bar{H}-\bar{\mathcal{N}}$ correlations}
\label{correlSect}

From the preceding sections it is clear that there does not need to be any correlation between the scalar field flux crossing $\bar{H}$ and that crossing $\bar{\mathcal{N}}$. These fluxes are 
actually free data. Any correlations will result from appropriate initial configurations of the fields.
In this final example we consider a physically interesting case where such a  correlation  exists. 

Consider  quadratic affine final data (Appendix \ref{daff}) on $\bar{H} = \{(v, 0): v_i < v < v_f\}$:
\begin{align}
\Phi_\ell^{(0)} = a_0 e^V + a_1e^{2V}  + a_2 e^{3V} \label{quadratic}
\end{align}
for $V = \nicefrac{v-v_f}{2R_o}$ along with similarly quadratic affine data on $\bar{\mathcal{N}}$:
\begin{align}
\Phi_{N_f} = \Phi_{N_f}^{(0)} +  \rho \Phi_{N_f}^{(1)} + \frac{\rho^2}{2} \Phi_{N_f}^{(2)} . 
\end{align}
A priori these are uncorrelated but let us restrict the initial configuration so that $\Phi_{N}^{(n)}(v_i) = 0$. That is, there is no  $\Phi_N$ flux through $v=v_i$. 
%We restrict our attention to order $\epsilon^1$ and so scalar field propagating in a Schwarzschild background.

Then the process to apply these conditions is, given the free final data on $\bar{H}$:
\begin{enumerate}[i)]
\item Solve for the $\Phi_N^{(n)}$ from (\ref{PN1}), (\ref{PN2}) and (\ref{PN3}).
\item Solve $\Phi_N^{(n)} (v_i) = 0$ to find the $\Phi_{N_f}^{(n)}$ in terms of the $a_n$. These are linear equations
and so the solution is straightforward. 
\item Substitute the resulting expressions for $\Phi_N^{(n)}$ into results from the previous sections to find all other quantities. 
\end{enumerate}
These calculations are straightforward but quite messy. Here we only present the final results for $\Phi_{N_f}$:
%
%
%
%Then these can be solved for $\Phi_{N_f}^{(n)}$ so that $\Phi_N$ vanishes at $V=V_i$  This amounts
%to three constraining equations on a set of six variables $\{a_o,a_1,a_2,\Phi_{N_f}^{(0)}, \Phi_{N_f}^{(1)},\Phi_{N_f}^{(2)} \}$. In fact this equations are linear and so it is not surprising that the 
%$\Phi^{(n)}_{N_f}$ can be written as linear combinations of the $a_n$. These calculations are straightforward but each expression is quite complicated and not particularly enlightening. As such we just
%present the final results. If the flux across $\bar{\mathcal{N}}$ is vanishing then:
\begin{align}
\Phi_{N_f}^{(0)}  \approx &( 1\! - e^{2V_i}) a_0 + \frac{2a_1(1\!-e^{3V_i})}{3}  +  \frac{a_2 (1\!-e^{4V_i})}{2} \\
\Phi_{N_f}^{(1)}  \approx & \frac{2a_0 (e^{2V_i} - e^{3V_i})}{R_o} + \frac{a_1 (1 + 8 e^{3V_i} - 9 e^{4V_i} )}{6R_o} \\
 & + \frac{a_2 (1 + 5 e^{4 V_i} - 6 e^{5V_i} )}{5R_o} \nonumber \\
 \Phi_{N_f}^{(2)}  \approx &  -\frac{a_0 (1+14 e^{2V_i} - 48 e^{3V_i} +33 e^{4V_i} ) }{4R_o^2} \\
 & -\frac{a_1 (1 + 35 e^{3V_i} - 135 e^{4V_i} + 99 e^{5V_i} ) }{15R_o^2} \nonumber \\
 &  +\frac{a_2 (1 -35 e^{4V_i} + 144 e^{5V_i} -110 e^{6V_i} ) }{20R_o^2}
 \nonumber
\end{align}
where $V_i = V(v_i)$. If $V_i$  is sufficiently negative that we can neglect  the exponential terms:
\begin{align}
\Phi_{N_f}^{(0)}  \approx &\;  a_0 + \frac{2a_1}{3}  +  \frac{a_2 }{2} \\
\Phi_{N_f}^{(1)}  \approx & \frac{a_1}{6R_o}  + \frac{a_2}{5R_o} \nonumber \\
 \Phi_{N_f}^{(2)}  \approx &  -\frac{a_0 }{4R_o^2} +
  -\frac{a_1 }{15R_o^2}
   + \frac{a_2  }{20R_o^2} \; . 
 \nonumber
\end{align}
In either case the flux through $\bar{H}$ fully determines the flux through $\bar{\mathcal{N}}$. The constraint at $v_i$ is sufficient to determine the Taylor expansion of the flux through $\bar{\mathcal{N}}$ relative to the expansion of the 
flux through $\bar{H}$. Though we only did this to second order in $\rho / v$ we expect the same process to fix the expansions to 
arbitrary order.

\section{Discussion}
\label{sec:discussion}
In this paper we have begun building a formalism that constructs spacetime in the causal past of a horizon $\bar{H}$
and an intersecting ingoing null surface $\bar{{\cal N}}$ using final data on those surfaces. It can be thought of as a specialized
characteristic initial value formulation and is particularly closely related to that developed in \cite{Winicour:2013gha}.
Our main interest has been to use the formalism to better understand the relationship between
horizon dynamics and off-horizon fluxes. So far we have restricted our attention to spherical symmetry and so included matter fields to drive the  dynamics. 

One of the features of characteristic initial value problems is that they isolate free data that may be specified on each of the 
initial surfaces. Hence it is no surprise that the corresponding data in our formalism is also free and uncorrelated. We considered
two types of data: inward flowing null matter and massless scalar fields. 

For the inward-flowing null matter, data on the horizon 
actually determines the entire spacetime running backwards along the ingoing null geodesics that cross $\bar{H}$. Physically this makes
sense. This is the only flow of matter and so there is nothing else to contribute to the dynamics. 

More interesting are the massless scalar field spacetimes. In that case, matter can flow both inwards and outwards and further 
inward moving radiation can scatter outwards and vice versa.  For the weak field near-horizon regime that we studied most
closely, the free final data is the scalar field flux through $\bar{H}$ and $\bar{\mathcal{N}}$ along with the value of $R$ at their
intersection. Hence, as noted, these fluxes are uncorrelated. However we also considered the case where there was no 
initial flux of scalar field  travelling ``up'' the horizon. In this case the coefficients of the Taylor expansion of the inward flux
on $\bar{H}$ fully determined those on $\bar{\mathcal{N}}$ (though in a fairly complicated way). 
This constraint is physically reasonable: one would expect the dominant matter fields close to a black hole horizon to be
infalling as opposed to travelling (almost) parallel to the horizon. It is hard to imagine a mechanism for generating strong parallel fluxes. 

While we have so far worked in spherical symmetry the current work still suggests ways to think about the
 horizon-$\mathscr{I}^+$ correlation problem for general spacetimes. 
For a dynamic non-spherical vacuum spacetime, gravitational wave fluxes will be the analogue of the
scalar field fluxes of this paper and almost certainly they will also be free data. Then any correlations will necessarily result
from special initial configurations. However as in our example these may not need to be very exotic. It may be sufficient to 
eliminate strong outward-travelling near horizon fluxes. 
In future works we will examine these more general cases in detail.

\acknowledgments

This work was supported by NSERC Grants 2013-261429 and 2018-04873. We are thankful to Jeff Winicour for discussions on characteristic evolution during the 2017
Atlantic General Relativity Workshop and Conference at Memorial University. IB would like to thank Abhay Ashtekar, Jos\'e-Luis Jaramillo and Badri Krishnan 
 for discussions during the 2018 ``Focus Session on Dynamical Horizons, Binary Coalescences, Simulations and Waveforms''  at Penn State. Bradley Howell
 pointed out a correction to our general integration scheme which is now incorporated in Sections \ref{flux:onh} and \ref{sec:int}. 

\appendix
\section{Causal past of $\bar{H} \cup \bar{\cal{N}}$}
\label{causalityApp}

In this appendix we consider how the general integration scheme for the scalar field spacetimes of 
 Section \ref{sec:matter modelsII} ``knows'' how to stay within the past domain of dependence of $\bar{H} \cup \bar{\mathcal{N}}$. 
 \begin{figure}[h]
\begin{center}
\includegraphics[scale=1]{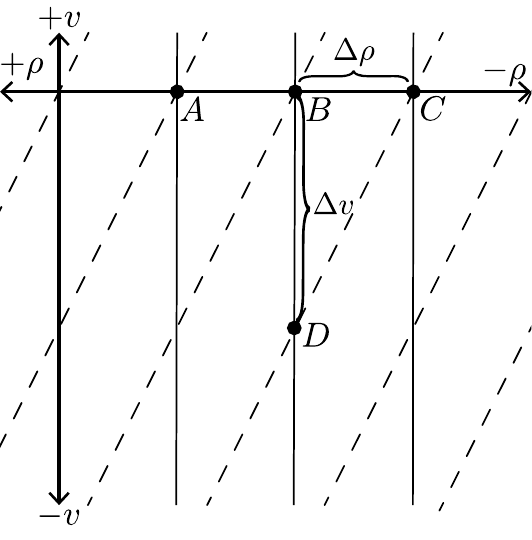} 
\caption{Causality restrictions on $\Delta v$: the CFL condition restricts the choice of $\Delta v$ to ensure that attempted numerical evolutions respect causality. In this
figure the $\rho$ and $v$ coordinates are drawn to be perpendicular to clarify the connection with the usual advection equation: to compare to other diagrams rotate about $45^\circ$ clockwise and skew so coordinate curves are no longer perpendicular. The dashed lines are null and have slope
$C$ in this coordinate system. If data at points $A$, $B$ and $C$ are used to determine $\Phi_{N,\rho}$ then the size of the discrete $v$-evolution is limited to lie 
inside the null line from point $C$.  The largest $\Delta v$ allowed by the restriction evolves to $D$. }
\label{causality}
\end{center}
\end{figure}

First, it is clear how the process develops spacetime up to the bottom left-hand null boundary ($v=v_i$) of the past domain of dependence. The bottom right-hand boundary is a little more complicated but follows from the advection form of the $\Phi_{N,v}$ equation (\ref{adN}). Details will depend on the exact numerical scheme 
but the general picture is as follows. 

Assume that we have discretized the problem so that we are working 
at points $(v_j, \rho_k)$. Then in using (\ref{adN}) to move from a surface $v_i$ to $v_{i-1}$, the 
Courant-Friedrichs-Lewy (CFL) condition (common to many hyperbolic equations) tells us that the maximum allowed $\Delta v$ is 
\begin{align}
\Delta v < \frac{\Delta \rho}{C} \, ,
\end{align}
where $\Delta \rho$ is the coordinate separation of the points that we are using to calculate the right-hand side of (\ref{adN}). 

Then, as shown in FIG.~\ref{causality}, the 
discretization progressively loses points of the bottom right of the diagram: they are outside of the domain of dependence of the individual points being used to determine them. 
For example if we are using  
a centred derivative so that 
\begin{align}
\Phi_{N,\rho} \approx  \frac{  \Phi_N(v_{j},\rho_{k+1} )- \Phi_N(v_{j},\rho_{k-1})  }{2 \Delta \rho}% \frac{\Phi_N(v_{j},\rho_{k+1} )- \Phi_N(v_{j},\rho_{k-1}) }{2 \Delta \rho} 
\end{align}
then we  need adjacent points as shown in FIG.~\ref{causality}. 

\begin{figure}[h]
\begin{center}
\includegraphics[scale=1]{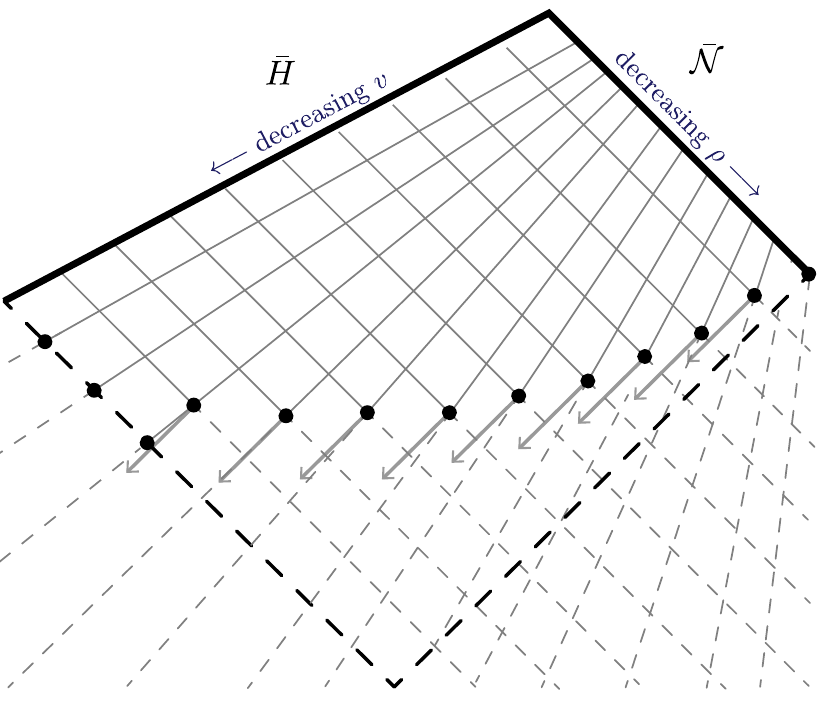} 
\caption{A cartoon showing the CFL-limited past domain of dependence of $\bar{H} \cup \bar{\mathcal{N}}$. Null lines are now drawn at $45^\circ$ so the analytic past domain of 
dependence is bound by the heavy dashed null lines running back from the ends of $\bar{H}$ and $\bar{\mathcal{N}}$. A (very coarse) discretization is depicted by the gray 
lines and the region that cannot be determined with dashed lines. The boundary points of that region are heavy dots.  }
\label{causality_2}
\end{center}
\end{figure}
The lower-right causal boundary of FIG.~\ref{dod} is then enforced by a combination of the endpoints of $\bar{\mathcal{N}}$ and the CFL condition as shown in 
FiG.~\ref{causality_2}. Points are progressively lost as they require greater than the maximum allowed $\Delta v$.
 The numerical past-domain of dependence necessarily lies inside the analytic domain. The coarseness of the discretization in the figure dramatizes
the effect: a finer discretization would keep the domains closer. 

%For any particular scheme of numerical integration there may be other restrictions on accuracy and the domain however the 

\section{Affine derivatives and final data}
\label{daff}
The off-horizon $\rho$-coordinate in our coordinate system is affine while $v$ is not. However, as seen in the main text, when considering the final data on $\bar{H}$ it is more natural to work relative to an affine parameter. This is somewhat 
complicated because $\Phi_\ell$ and $\Phi_N$ are respectively linearly dependent on $\ell$ and $N$ and the scaling of those vectors is also tied to coordinates via (\ref{E1}), (\ref{E2}) and (\ref{crel1}). 
In this appendix we will discuss the affine parameterization of the horizon and the associated affine derivatives for various quantities. 

Restricting our attention to an isolated horizon $\bar{H}$ with $\kappa = \frac{1}{2R_o}$, consider a reparameterization 
\begin{align} 
\tilde{v} = \tilde{v} (v) \, . 
\end{align}
Then
\begin{align}
\frac{\partial}{\partial v}  = \frac{\mbox{d} \tilde{v}}{\mbox{d} v} \frac{\partial}{\partial \tilde{v}}
\end{align}
and so 
\begin{align}
 \ell  = e^V  \tilde{\ell} \; \; \mbox{and} \; N = e^{-V} \tilde{N}
\end{align}
where we have defined $V$ so that
$ \displaystyle
e^V=\frac{\mbox{d} \tilde{v}}{\mbox{d} v}   \, . 
$
Hence
\begin{align}
\tilde{\kappa} = - \tilde{N}_b \tilde{\ell}^a \nabla_a \tilde{\ell}^b = e^{-V} \left(\kappa - \frac{\mbox{d} V}{\mbox{d} v}  \right)  \,  
\end{align}
and so for an affine parameterization ($ 
\kappa=\partial_v V   
$):
\begin{align}
 e^V  = \exp \left( \frac{v-v_f}{2R_o} \right)
 \end{align}
 for some $v_f$ and 
\begin{align}
\tilde{v} - \tilde{v}_o  = 2 R_o e^V % \; \; \mbox{for} \; V = \frac{v-v_o}{2R_o}
\end{align}
for some $\tilde{v}_o$. The $v_f$ freedom corresponds to the freedom to rescale an affine parameterization by a constant multiple while the $\tilde{v}_o$ is the freedom to set the zero of $\tilde{v}$ wherever you like. 

Now consider derivatives with respect to this affine parameter. For a regular scalar field
\begin{align}
\frac{\mbox{d} f}{\mbox{d} \tilde{v}} =   e^{-V} \frac{\mbox{d}f}{\mbox{d} v} \, . 
\end{align}
However in this paper we are often interested in scalar quantities that are defined with respect to the null vectors:
\begin{align}
\Phi^{(0)}_{\ell} = e^{V} \Phi^{(0)}_{\tilde{\ell}}  \; \; \mbox{and} \; \Phi^{(0)}_{N} = e^{-V} \Phi^{(0)}_{\tilde{N}} \, . 
\end{align}
Then
\begin{align}
 \frac{\mbox{d}  \Phi^{(0)}_{\tilde{\ell}}}{\mbox{d} \tilde{v}}\!  &= \! e^{-V} \!  \frac{\mbox{d}}{\mbox{d} v} \left(e^{-V} \Phi^{(0)}_\ell \! \right) \!  = \! e^{-2V} \!  \! \left(\!  \frac{\mbox{d} \Phi^{(0)}_\ell}{\mbox{d} v} \!  - \kappa \Phi^{(0)}_\ell \! \! \right) \\
\frac{\mbox{d}  \Phi^{(0)}_{\!\tilde{N}}}{\mbox{d} \tilde{v}}\!  &= e^{-V} \!   \frac{\mbox{d}}{\mbox{d} v} \left(e^{V} \Phi^{(0)}_N \right) =  \frac{\mbox{d} \Phi^{(0)}_N}{\mbox{d} v} \! + \kappa \Phi^{(0)}_N \, . 
\end{align}
That is these quantities are affinely constant if 
\begin{align}
\Phi_\ell  = e^{V} \Phi^{(0)}_{\ell_f}  \; \; \mbox{and} \;
\Phi_N  = e^{-V} \Phi^{(0)}_{N_f} 
\end{align}
for some constants $\Phi^{(0)}_{\ell_f}$ and $\Phi^{(0)}_{N_f}$. 

In the main text we write this affine derivative on $\bar{H}$ as $D_v$ with its exact form depending on the $\ell$ or $N$ 
dependence of the quantity being differentiated.

Finally at (\ref{quadratic}) we consider a $\Phi_\ell$ that is ``affinely quadratic''. By this we mean that:
\begin{align}
 \Phi_{\tilde{\ell}} &= A_o + A_1 \tilde{v}  + A_2 \tilde{v} ^2  \nonumber \\
& \mspace{70mu} \Updownarrow \nonumber \\
\Phi_\ell &= a_o e^{V} + a_1 e^{2V} +  a_2  e^{3V}  \, ,
\end{align}
where for simplicity we have set $\tilde{v}_o$ to zero (so that  $v=0$ is $\tilde{V}=2R_o$) and absorbed the extra $2R_o$s into the $a_n$.

\end{document}